\documentclass[11pt]{article}
\pdfoutput=1
\usepackage{soul}
\usepackage{jheppub}
\usepackage{amsfonts}
\usepackage{amsmath}
\allowdisplaybreaks[4]         
\usepackage{amssymb}   
\usepackage{euscript}       
\usepackage{xcolor}          
\usepackage{tensor}     
\usepackage{caption}
\usepackage{subcaption}   
\usepackage{graphicx}

\newcommand{\req}[1]{(\ref{#1})} 

\newcommand{\bea}{\begin{eqnarray}}
\newcommand{\eea}{\end{eqnarray}}
\newcommand{\ba}{\begin{eqnarray}}
\usepackage{braket}
\newcommand{\ea}{\end{eqnarray}}

\newcommand{\beq}{\begin{equation}}
\newcommand{\eeq}{\end{equation} }
\newcommand{\beqa}{\begin{eqnarray}}

\newcommand{\eeqa}{\end{eqnarray}}
\newcommand{\beqar}{\begin{eqnarray*}}
\newcommand{\eeqar}{\end{eqnarray*}}

\newcommand{\be}{\begin{equation}}
\newcommand{\ee}{\end{equation}}
\newcommand{\diff}{\mathrm{d}}

\newcommand{\dv}{\\\notag &}
\newcommand{\dvv}{\right.\\\notag &\left.}

\renewcommand{\req}[1]{(\ref{#1})}

\allowdisplaybreaks

\title{Near-horizon geometries and black hole thermodynamics in higher-derivative AdS$_5$ supergravity}

\author[a,b]{Pablo A. Cano,}
\author[a]{Marina David}

\affiliation[a]{Instituut voor Theoretische Fysica, KU Leuven.
Celestijnenlaan 200D, B-3001 Leuven, Belgium \vspace{0.1cm}}
\affiliation[b]{Departament de F\'isica Qu\`antica i Astrof\'isica, Institut de Ci\`encies del Cosmos Universitat de Barcelona, Mart\'i i Franqu\`es 1, E-08028 Barcelona, Spain}

\emailAdd{pablo.cano@icc.ub.edu}
\emailAdd{marina.david@kuleuven.be}

\date{\today}
\abstract{
Higher-derivative corrections in the AdS/CFT correspondence allow us to capture finer details of the dual CFT and to explore the holographic dictionary beyond the infinite $N$ and strong coupling limits. Following an effective field theory approach, we investigate extremal AdS black hole solutions in five-dimensional supergravity with higher-derivative corrections. We provide a general analysis of near-horizon geometries of rotating extremal black holes and show how to obtain their corresponding charges and chemical potentials. We discuss the near-horizon solutions of the two-derivative theory, which we write using a novel parametrization that eases our computation of the higher-derivative corrections. The charges and thermodynamic properties of the black hole are computed while clarifying the ambiguities in their definitions. The charges and potentials turn out to satisfy a near-horizon version of the first law of thermodynamics whose interpretation we make clear.  In the supersymmetric case, the results are shown to match the field theory prediction as well as previous results obtained from the on-shell action.  }

\begin{document} 
\maketitle
\flushbottom

\newpage
\allowdisplaybreaks
\section{Introduction}
\label{sec:Introduction}

The study of black holes in AdS spacetimes has driven a great deal of investigation in fundamental physics, as the AdS/CFT correspondence \cite{Maldacena:1997re, Witten:1998qj, Gubser:1998bc} has led to significant progress in our understanding of both gravity and gauge field theories.
In particular, precision holography, which aims to investigate the duality between gravity and conformal field theories beyond leading order in the large $N$ and strong coupling regime, is currently the focus of intensive research. In the gravity side, this involves studying corrections to AdS spacetimes, such as higher derivative corrections \cite{Bobev:2020zov,Bobev:2020egg,Bobev:2021qxx, Bobev:2021oku,Bobev:2022bjm, Cassani:2022lrk, Cassani:2023vsa, Cano:2023dyg} or 1-loop effects, like those of logarithmic type \cite{Bhattacharyya:2012ye, David:2019lhr, GonzalezLezcano:2020yeb, David:2021eoq, David:2021qaa,Karan:2022dfy, Bobev:2023dwx}. 
By their own nature, the study of these corrections is a challenging endeavour. In this work, we focus on higher derivative corrections to black hole solutions of minimal gauged five-dimensional supergravity, with a special focus (but not only) on supersymmetric solutions.

The black hole solutions in this theory are in general characterized by the mass, electric charge and two different angular momenta. As usual in the case of AdS, the supersymmetric solutions must have rotation, which increases the complexity of their analysis. The simplest of those solutions was first written by Gutowski and Reall \cite{Gutowski:2004ez}, describing a black hole with electric charge and two equal angular momenta.  More general solutions of this theory --- with arbitrary mass, charge and distinct angular momenta, and containing the most general supersymmetric solution as a particular case --- were found in \cite{Chong:2005hr}. Our focus will be on studying corrections to this solution, which we call the CCLP solution. 

There has been recent work in understanding higher derivative corrections in five-dimensional supergravity. The starting point involves applying the off-shell formalism of $\mathcal{N}=2$ gauged supergravity \cite{Lauria:2020rhc} with all possible four-derivative invariants \cite{Zucker:1999ej, deWit:2006gn,Hanaki:2006pj,Cremonini:2008tw,Bergshoeff:2011xn,Coomans:2012cf,Ozkan:2013nwa,Baggio:2014hua, Butter:2014xxa, Bobev:2021qxx, Cassani:2022lrk,Liu:2022sew,Gold:2023dfe,Gold:2023ykx,Gold:2023ymc}. Integrating out the auxiliary degrees of freedom leads to an explicit action for the metric and gauge field \cite{Bobev:2021qxx, Liu:2022sew, Cassani:2022lrk}, which can be further simplified by implementing field redefinitions \cite{Cassani:2022lrk,Gold:2023ymc}. From the point of view of holography, such theory is now dual to a superconformal field theory with unequal central charges $\mathfrak{a}\neq \mathfrak{c}$, the difference $(\mathfrak{a}-\mathfrak{c})$ being proportional to the higher-derivative couplings \cite{Hofman:2008ar}. 
 
Refs.~\cite{Bobev:2022bjm, Cassani:2022lrk} have recently been able to compute the corrections to the thermodynamic quantities of the CCLP black hole\footnote{See also \cite{Gold:2023ymc} for a similar analysis.}, showing that, in the supersymmetric case, the result precisely agrees with the corresponding superconformal index of the dual CFT, which contains corrections controlled by difference of central charges $(\mathfrak{a}-\mathfrak{c})$. These references made use of the on-shell action in order to compute the free energy in the grand canonical ensemble, from which one derive the subsequent thermodynamic quantities, such as energy, charges and entropy.  A key aspect of their computation is that, in order to obtain the free energy at first order in the higher-derivative corrections, it suffices to evaluate the on-shell action on the two-derivative solution, following the Reall and Santos method \cite{Reall:2019sah} --- see also \cite{Melo:2020amq,Ma:2023qqj,Hu:2023gru}. Thus, their approach avoids the complicated task of solving the equations of motion. 

Here we take a different approach in order to study the corrections to the CCLP solution. We consider extremal (but not necessarily supersymmetric) black holes and we study their near-horizon geometries. Due to the enhanced symmetry of the near-horizon region, solving the equations of motion is much easier than for the full solution. Thus, our goal is to compute the corrections to these solutions and to their thermodynamic quantities by studying their near-horizon geometries.

There are several reasons that motivate us to pursue this goal. First, even though the results of \cite{Bobev:2022bjm, Cassani:2022lrk} perfectly agree with field theory expectations, the Reall-Santos approach depends on carefully dealing with boundary terms in the on-shell action, which is even more subtle in the case of AdS spacetimes \cite{Hu:2023gru}. Thus, we consider it wise to derive those results by a different method. In addition, the on-shell action approach does not tell us much about the interpretation of the thermodynamic charges derived from it. In fact, \cite{Cassani:2023vsa}, which studied supersymmetric near-horizon geometries with equal angular momenta, already observed a discrepancy between the thermodynamic electric charge and the standard definition of charge understood as a surface integral. Here we further analyze this issue for general near-horizon geometries.  
Another reason for following the near-horizon approach is that it presumably allows for an easier generalization to subleading corrections with respect to the on-shell action method, as we argue in Section~\ref{sec:conclusions}. Finally, obtaining a solution of the equations of motion (in our case, the near-horizon geometry) gives us much more information than just the thermodynamic properties and can be useful for other purposes, like \textit{e.g.} the study of perturbations \cite{Castro:2021wzn}.

The study of near-horizon extremal geometries does present some challenges. 
Unlike the case of spherically symmetric black holes, where one can elegantly reduce the problem to an algebraic system of equations by using the entropy function formalism \cite{Sen:2005wa, Sen:2005iz,Dabholkar:2006tb,Sen:2007qy}, for rotating black holes the story is more involved \cite{Astefanesei:2006dd,Morales:2006gm,LopesCardoso:2007hen,Kunduri:2007vf,Kunduri:2007qy,Cano:2019ozf,Cano:2023dyg}. In the case of five-dimensional solutions, black holes with two equal angular momenta are of cohomogeneity 1 and have an enhanced symmetry group that leads to a drastic simplification of the near-horizon solutions \cite{Gutowski:2004ez,Cassani:2023vsa}. However, when the two angular momenta are different, the solutions take a more complicated form and we are required to solve a very intricate system of differential equations. 
On top of this, there is the question of identifying the various charges and potentials of the black hole from the near-horizon region. By following the approaches of \cite{Bazanski:1990qd,Kastor:2008xb, Kastor:2009wy, Ortin:2021ade,Cano:2023dyg,Cassani:2023vsa}, we will compute the charges (electric charge and angular momenta) by using Komar integrals \cite{Komar:1958wp} that are independent of the surface of integration and hence can be evaluated at the horizon. The potentials (electrostatic potential and angular velocities of the horizon) are on the other hand defined as differences of certain quantities between the horizon and infinity, and hence cannot be determined from the near-horizon region alone. Nevertheless, one can identify unambiguously certain quantities that we will refer to as \textit{near-horizon potentials.} It turns out that the entropy, charges and these potentials satisfy a near-horizon version of the first law of thermodynamics, which was already observed (in the case of equal angular momenta supersymmetric solutions) in  \cite{Cassani:2023vsa}. Here we show that this law holds for general near-horizon extremal geometries (including higher-derivative corrections) and we provide an explanation for it that allows us to interpret the near-horizon potentials.\footnote{See \cite{Hajian:2013lna, Hajian:2014twa} for a more general investigation of the laws of black hole thermodynamics at the near-horizon region.}

In the supersymmetric case, our results reproduce those of \cite{Bobev:2022bjm, Cassani:2022lrk} up to an ambiguity in the definition of the electric charge. On the other hand, for most of our analysis we do not restrict ourselves to supersymmetric solutions and we consider general extremal black holes with arbitrary charge and different angular momenta.

This work is organized as follows. In Section~\ref{sec:AdS5sugra}, we review the details of the five-dimensional supergravity action with higher derivative corrections. We discuss how to compute the conserved charges and we review the CCLP solution of the two-derivative theory. 
Section~\ref{sec:NHG} is devoted to a general analysis of near-horizon geometries. We introduce an appropriate ansatz and show how to identify the angular coordinates and near-horizon potentials in full generality. We then move on to Section~\ref{sec:2der} where we obtain the near-horizon geometry of the extremal CCLP black holes. We introduce a novel parametrization that allows us to write the solution in a fully explicit and compact form, and we derive and study the various thermodynamic quantities and relations. 
The analysis of the four-derivative corrections to the solution comprises Section~\ref{sec:higher} while we focus on the corrected thermodynamics in Section~\ref{sec:thermo}.  We conclude with some remarks and open problems in Section~\ref{sec:conclusions}. In Appendix \ref{app:solution} we provide some explicit formulas for the near-horizon solutions that are too long to be displayed in the main text. We include as well a Mathematica notebook with our (even longer) complete results. 

\section{Review of higher-derivative AdS$_5$ supergravity}
\label{sec:AdS5sugra}
\subsection{Action  }
In this section we briefly review the five-dimensional gauged supergravity theory of interest. We focus on preparing ourselves to study the four-derivative corrections of the theory and the corresponding Noether surface charges. After a careful series of field redefinitions starting from the off-shell formalism  \cite{Bobev:2021qxx, Cassani:2022lrk}, the result of the supergravity action is relatively compact
\begin{equation}\label{eq:action}
	\begin{aligned}
		S\,=&\,\frac{1}{16\pi G_5}\int \diff^5x \,\sqrt{|g|} \left\{ c_0 R+12c_1g^2-\frac{c_2}{4}F^2-\frac{c_3}{12\sqrt{3}}\epsilon^{\mu\nu\rho\sigma\lambda}F_{\mu\nu}F_{\rho\sigma} A_{\lambda}\right.\\[2mm]
		&\left.\,+\,\lambda_1 \alpha g^2 \left[{\cal X}_{\text{GB}}-\frac{1}{2}C_{\mu\nu\rho\sigma}F^{\mu\nu}F^{\rho\sigma}+\frac{1}{8}F^4-\frac{1}{2\sqrt{3}}\epsilon^{\mu\nu\rho\sigma\lambda}R_{\mu\nu\alpha\beta}R_{\rho\sigma}{}^{\alpha\beta} A_{\lambda}\right]\right\}\, ,
	\end{aligned}
\end{equation}
where $g$ is equal to the inverse length of AdS and the Gauss-Bonnet invariant $\mathcal{X}_{\text{GB}}$, Weyl tensor $C_{\mu\nu\rho\sigma}$ and the Maxwell field invariants $F^2$ and $F^4$ are given by
\begin{subequations}
\begin{align}
	{\cal X}_{\text{GB}}&=R_{\mu\nu\rho\sigma}R^{\mu\nu\rho\sigma}-4 R_{\mu\nu}R^{\mu\nu}+R^2\, ,
	\\
	C_{\mu\nu\rho\sigma}&=R_{\mu\nu\rho\sigma}-\frac{2}{3}\left(R_{\mu[\rho}g_{\sigma]\nu}+R_{\nu[\sigma}g_{\rho]\mu}\right)+\frac{1}{6}Rg_{\mu[\rho}g_{\sigma]\nu}\, ,\\
	F^2&= F_{\mu\nu}F^{\mu\nu}\, ,\\
	F^4&= F_{\mu\nu}F^{\nu\rho}F_{\rho\sigma}F^{\sigma\mu}\, .
\end{align}
\end{subequations}
The various constants $c_i$ are defined as ${c}_{i}=1+\alpha g^2 {\delta c}_{i}$ with
\begin{equation}\label{cdef}
	{\delta c}_{0}=4\lambda_2\, , \hspace{.5cm}{\delta c}_{1}=-10\lambda_1+4\lambda_2 \,, \hspace{.5cm} { \delta c}_{2}=4\lambda_1+4\lambda_2 \,, \hspace{.5cm}{\delta 
		 c}_{3}=-12\lambda_1+4\lambda_2 \,,
\end{equation}
where $\lambda_1$ and $\lambda_2$ are two free coupling constants that control the corrections. The action is written in a way such that the first line in \eqref{eq:action} is the usual two-derivative contribution when $\alpha$ is set to zero.
We note that the effect of $\lambda_{2}$ is just to renormalize the coupling constants and fields of the two-derivative action. In particular, since it changes the normalization of the Einstein-Hilbert term, it is natural to introduce an effective Newton's constant, 
\begin{equation}\label{Geff}
G_{\rm eff}=\frac{G_5}{1+4\alpha g^2 \lambda_{2}}\, ,
\end{equation} 
which will naturally appear in some of our results. From the holographic perspective, the $\lambda_{i}$ couplings modify the CFT central charges $\mathfrak{a}$ and $\mathfrak{c}$, and in particular they break the degeneracy between them --- this is a generic effect of higher-derivative terms \cite{Hofman:2008ar,Buchel:2009sk,Myers:2010jv,Bueno:2018xqc,Cano:2022ord}. In fact, we have \cite{Bobev:2021qxx, Cassani:2022lrk}
\begin{equation}\label{acdef}
\mathfrak{a}=\frac{\pi}{8 g^3G_{\rm eff}}\, ,\qquad \mathfrak{a}-\mathfrak{c}=-\frac{\pi\alpha\lambda_{1}}{g G_{\rm eff}}\, .
\end{equation}
\subsection{Equations of motion and conserved quantities  }

From the action \eqref{eq:action}, we can derive the equations of motion $\mathcal{E}_{\mu\nu}$ and $\mathcal{E}_{\mu}$ by varying the metric as well as the gauge potential, respectively. This gives us
\begin{eqnarray} \label{eq:eoms}
	\mathcal{E}_{\mu\nu} &=& -\frac{1}{2}g_{\mu\nu}{\cal L}'+P^{\alpha\beta\gamma}{}_{(\mu|}R_{\alpha\beta\gamma |\nu)}-2\nabla^{\alpha}\nabla^{\beta} (P_{\beta (\mu\nu)\alpha}+\Pi_{\beta (\mu\nu)\alpha})-\frac{1}{2}\mathcal{F}_{(\mu|\rho}F_{|\nu)}{}^{\rho}\, \label{EinsteinEqCorrected},\\[2mm]
	\mathcal{E}_{\mu} &=& -\nabla_{\mu}\mathcal{F}^{\mu\nu}+\frac{ c_3}{4\sqrt{3}}\epsilon^{\nu\alpha\beta\gamma\delta}F_{\alpha\beta}F_{\gamma\delta}+\frac{\lambda_1\,\alpha}{2\sqrt{3}}\epsilon^{,nu\alpha\beta\gamma\delta}R_{\alpha\beta\rho\sigma}R_{\gamma\delta}{}^{\rho\sigma}\, .
\end{eqnarray}
where
\begin{subequations}
	\begin{align}
		P_{\mu\nu\rho\sigma}=&\frac{\partial {\cal L}'}{\partial R_{\mu\nu\rho\sigma}}=\, {  c}_{0}g_{\mu[\rho}g_{\sigma]\nu}+\lambda_1\,\alpha\left[2R_{\mu\nu\rho\sigma}-4\left(R_{\mu[\rho}g_{\sigma]\nu}-R_{\nu[\rho}g_{\sigma]\mu}\right)+2 g_{\mu[\rho}g_{\sigma]\nu} R \right. \nonumber\\[1mm]
		&\left.-\frac{1}{2}F_{\mu\nu}F_{\rho\sigma}-\frac{1}{12}g_{\mu[\rho}g_{\sigma]\nu} F^2+\frac{1}{3}\left(F_{\mu\alpha}F_{[\rho}{}^{\alpha}g_{\sigma]\nu}-F_{\nu\alpha}F_{[\rho}{}^{\alpha}g_{\sigma]\mu}\right)\right]\, ,\\[1mm]
		\Pi^{\mu\nu\rho\sigma} =&-\frac{\lambda_1\,\alpha}{\sqrt{3}}\epsilon^{\mu\nu\alpha\beta\gamma}R_{\alpha\beta}{}^{\rho\sigma}A_{\gamma}\, ,
		\\
	\mathcal{F}^{\mu\nu}=&-2\frac{\partial {\cal L}'}{\partial F_{\mu\nu}}=  c_2 F^{\mu\nu}-2\lambda_1\,\alpha \left(-C^{\mu\nu\rho\sigma}F_{\rho\sigma}+\frac{1}{2}F^{\mu\rho}F_{\rho\sigma}F^{\nu\sigma}\right)\,,
	\end{align}
\end{subequations}
and $\mathcal{L}'=\mathcal{L}-\mathcal{L}_{\rm CS}$ is the Lagrangian without the Chern-Simons terms.\footnote{We also point out that we define the Lagrangian without the $(16\pi G_{5})^{-1}$ factor, so $\mathcal{L}=c_0 R+\ldots$ . } 
It is manifest that these nonlinear coupled differential equations are highly complicated to solve. 
Nevertheless, our goal is to solve them for near-horizon geometries and to evaluate the charges from a first principles approach.
Here we review how obtain the conserved charges via the Noether current associated to spacetime symmetries and gauge transformations. 
For a thorough analysis of the conserved charges of the theory \req{eq:action} we refer to \cite{Cassani:2023vsa}. Here we quote their results and remark on some of the main subtleties. 

Due to the Chern-Simons terms, there are several notions of electric charge that one may consider \cite{Marolf:2000cb}. Here we will use the ``Page'' charge \cite{PhysRevD.28.2976}, which is given by 
\begin{align} \label{eq:Qint}
Q&=-\frac{1}{16\pi G_5}\int_{\Sigma}\left(\star {\cal F}-\frac{c_3}{\sqrt{3}}F\wedge A-\frac{2\lambda_1\alpha}{\sqrt{3}}\Omega_{\rm{CS}}\right)\,,
\end{align}
where $\Omega_{\rm{CS}}$ is the Lorentz-Chern-Simons three-form defined by
\begin{equation}\label{eq:lorentzchernsimons}
	\Omega_{\rm{CS}}=\diff \omega^{ab}\wedge \omega_{ab}-\frac{2}{3}\, \omega^{a}{}_{b}\wedge \omega^{b}{}_{c}\wedge \omega^{c}{}_a\,,
\end{equation}
and $\omega^{ab}=\,\omega^{ab}{}_{\mu}\,\diff x^{\mu}$ is the spin connection and the Latin indices are flat. This charge satisfies a Gauss law, so the result is independent of the choice of integration surface $\Sigma$, as long as this is a co-dimension 2 spacelike hypersurface homeomorphic to the sphere at infinity. Thus, it can be evaluated at the black hole horizon, $\Sigma=\mathcal{H}$. On the other hand, $Q$ is gauge and frame dependent, due to the gauge and Lorentz-Chern-Simons three-forms, respectively. The gauge ambiguity can be fixed by demanding regularity of the vector field at the horizon, but the frame ambiguity is more worrisome as there is no canonical choice of frame.  We revisit this point in detail in Section~\ref{subsec:Qamb}.

If the solution has a Killing vector $\xi_{\mu}$, one also finds the conserved current
\begin{equation}
\boldsymbol{J}_{\xi}=\diff \boldsymbol{Q}_{\xi}\, ,
\end{equation}
where $ \boldsymbol{Q}_{\xi}$ is the Noether charge three-form, given by 
\begin{equation}
 \boldsymbol{Q}_{\xi}= -\frac{1}{32\pi G_{5}}\boldsymbol {\epsilon}_{\mu\nu}\left[4\nabla_\sigma \mathcal P^{\mu\nu\sigma\rho}\xi_\rho - 2 \mathcal P^{\mu\nu\sigma\rho}\nabla_\sigma\xi_\rho - \left(\xi_\gamma A^\gamma\right)\left( \mathcal F^{\mu\nu} +\frac{c_3}{3\sqrt{3}}\epsilon^{\mu\nu\rho\sigma\lambda}A_\rho F_{\sigma\lambda} \right)\right]\, ,
\end{equation}
and where 
\begin{equation}
{\cal P}^{\mu\nu\rho\sigma}= \, P^{\mu\nu\rho\sigma}+\Pi^{\mu\nu\rho\sigma}-(P^{\mu[\nu\rho\sigma]}+\Pi^{\mu[\nu\rho\sigma]})\, ,
\end{equation}
where the antisymmetrized terms are there to guarantee that ${\cal P}^{\mu[\nu\rho\sigma]}=0$.\footnote{Those terms do not affect the equations of motion, but they are crucial in the computation of the Noether charge.} The integral of $\boldsymbol{J}_{\xi}$ over any Cauchy slice $\mathcal{C}$ then yields the conserved charge, and through the use of Stokes' theorem one can reduce this to an integral over the boundary, $\Sigma=\partial \mathcal{C}$,
\begin{equation}\label{eq:Jint}
J[\xi] =\int_{\Sigma}\boldsymbol{Q}_{\xi}\, .
\end{equation}
In this case, it is important that $\Sigma$ is the sphere at infinity, since $\diff \boldsymbol{Q}_{\xi}\neq 0$ and therefore we do not have a Gauss law. Hence, the conserved charges such as the total energy and angular momenta must in principle be computed at infinity. A way around this consists in defining a ``Noether-Komar'' charge three-form
\begin{equation}
\widetilde{\boldsymbol{Q}}_{\xi}=\boldsymbol{Q}_{\xi}-\boldsymbol{\Omega}\, ,
\end{equation}
where $\diff \boldsymbol{\Omega}=\boldsymbol{J}_{\xi}\big|_{\rm on-shell}$, so that we have $\diff\widetilde{\boldsymbol{Q}}_{\xi}=0$ on-shell. Finding such a three-form $\boldsymbol{\Omega}$ is always possible by noting that, on-shell, one has $\boldsymbol{J}_{\xi}=-\frac{1}{2}\star\boldsymbol{\xi}\mathcal{L}$. For more details about the construction of these Komar charges, we refer to \cite{Bazanski:1990qd,Kastor:2008xb, Kastor:2009wy,Ortin:2021ade,Cano:2023dyg,Cassani:2023vsa}. Now, the integral of $\widetilde{\boldsymbol{Q}}_{\xi}$ is independent of the surface of integration, and it computes the Noether charge as long as the integral of $\boldsymbol{\Omega}$ at infinity vanishes. Thus, one can use $\tilde{\boldsymbol{Q}}_{\xi}$ to evaluate the charges as an integral over the horizon. However, there is a final twist in this story. If one only has access to the near-horizon region, then one usually cannot fix the ambiguity in $\boldsymbol{\Omega}$ by demanding that its integral at infinity vanishes.  This is one of the reasons why it is not possible to compute the mass from the near-horizon region. Fortunately, in the case of the angular momenta, it is possible to fix this ambiguity. It turns out that, in a stationary spacetime, one can choose the  $\boldsymbol{\Omega}$ for the rotational Killing vectors in a way that its integral on any constant time surface\footnote{The time coordinate is taken as the coordinate associated to the time-like Killing vector.} vanishes \cite{Cassani:2023vsa}.  For a very explicit example of this we refer to \cite{Cano:2023dyg}. Thus, to make the long story short, it turns out one can actually compute the angular momenta by integrating  \req{eq:Jint} on the near-horizon region as long as we integrate on a constant time surface.

Finally, we are also interested in the entropy of the black hole, which is given by the Iyer-Wald formula \cite{Wald:1993nt,Iyer:1994ys}
\begin{equation}\label{eq:Waldentropy}
	S=-\frac{1}{8G_{5}}\int_{\cal H} \diff^3x\, \sqrt{\gamma}\,{\cal P}^{\mu\nu\rho\sigma}\, n_{\mu\nu}\, n_{\rho\sigma}\,,
\end{equation}
where $\gamma$ is the determinant of the induced three-dimensional horizon metric and $n_{\mu\nu}$ is the binormal to the horizon normalized via $n_{\mu\nu}n^{\mu\nu}=-2$. We note that this formula, just like those for the charge and the angular momenta, is gauge-dependent. It is possible to refine Wald's formalism in order to derive a gauge-invariant entropy formula \cite{Elgood:2020svt,Elgood:2020mdx,Elgood:2020nls}, but here we will fix this ambiguity by demanding regularity of the gauge field at the horizon. The validity of this approach is ultimately verified by checking that our results are compatible with the first law of thermodynamics and that they agree with the results of \cite{Bobev:2022bjm, Cassani:2022lrk}, obtained by different methods.

\subsection{Black holes: review of the CCLP solution  }
As we are interested in finding corrections to the five-dimensional AdS black hole with two distinct angular momenta and electric charge as was first constructed in \cite{Chong:2005hr}, we briefly review the solution, along with its thermodynamic quantities. The metric and gauge field are given by
\begin{equation}
	\begin{aligned}
		d s^2 = & -\frac{\Delta_\theta\left[\left(1+g^2 r^2\right) \rho^2 d t+2 q \nu\right] d t}{\Xi_a \Xi_b \rho^2}+\frac{2 q \nu \omega}{\rho^2}+\frac{\gamma}{\rho^4}\left(\frac{\Delta_\theta d t}{\Xi_a \Xi_b}-\omega\right)^2+\frac{\rho^2 d r^2}{\Delta_r}+\frac{\rho^2 d \theta^2}{\Delta_\theta} \\
		& +\frac{r^2+a^2}{\Xi_a} \sin ^2 \theta d \phi^2+\frac{r^2+b^2}{\Xi_b} \cos ^2 \theta d \psi^2, \\
		A= & \frac{\sqrt{3} q}{2\rho^2}\left(\frac{\Delta_\theta d t}{\Xi_a \Xi_b}-\omega\right), 
	\end{aligned}
\end{equation}	
where
\begin{equation} \label{eq:CCLPfunctions}
	\begin{aligned}
		\nu & =b \sin ^2 \theta d \phi+a \cos ^2 \theta d \psi, \quad &\Delta_r & =g^2\left(r^2+a^2\right)\left(r^2+b^2\right)\left(1+\frac{1}{g^2r^2}\right)+ \frac{q^2+2 a b q}{r^2}-2 m, \\
		\omega & =a \sin ^2 \theta \frac{d \phi}{\Xi_a}+b \cos ^2 \theta \frac{d \psi}{\Xi_b}, \quad &\Delta_\theta & = 1-a^2 g^2 \cos ^2 \theta-b^2 g^2 \sin ^2 \theta,
		\\
		\rho^2 & = r^2+a^2 \cos ^2 \theta+b^2 \sin ^2 \theta, \quad & \Xi_a & = 1-a^2 g^2,\\
		\gamma & = 2 m \rho^2-q^2+2 a b q g^2 \rho^2,
		\quad & \Xi_b & = 1-b^2 g^2.
	\end{aligned}
\end{equation}
The general non-extremal black hole is paramatrized by the four parameters $\{a,b,m,q\}$ associated to the angular momenta $J_1,J_2$, energy $E$ and electric charge $Q$. The conserved quantities $J_1,J_2$ and $Q$ may be found via the Komar integrals we introduced above (but specified for the two-derivative theory)
\begin{align} 
	\label{eq:CCLP-Ja}
	J_{1} &= \frac{1}{16\pi G_5}\int_{S^3} \star d \xi_{\phi} = \frac{\pi\left[2 a m+q b\left(1+a^2 g^2\right)\right]}{4 G_{5}\Xi_a^2 \Xi_b},
	\\ \label{eq:CCLP-Jb}
	J_{2} &= \frac{1}{16\pi G_5}\int_{S^3} \star d \xi_{\psi} = \frac{\pi\left[2 b m+q a\left(1+b^2 g^2\right)\right]}{4 G_{5}\Xi_b^2 \Xi_a},
	\\
	\label{eq:CCLP-Q}
	Q &= \frac{1}{16\pi G_{5}}\int_{S^3} \left( \star F - \frac{1}{\sqrt{3}} F \wedge A \right) = \frac{\sqrt{3} \pi q}{4 G_{5}\Xi_a \Xi_b},
\end{align}
where $\xi_{\phi}=- g_{\mu\phi} dx^{\mu}$ and $\xi_{\psi}=- g_{\mu\psi} dx^{\mu}$ are  the Killing vectors $-\partial_{\phi}$ and $-\partial_{\psi}$ expressed as 1-forms.

The Hawking temperature is derived by requiring appropriate periodic identifications in Euclidean time, which leads to
\begin{align} \label{eq:Tgeneral}
	T = \beta^{-1} &= \frac{r_{+}^4\left[\left(1+g^2\left(2 r_{+}^2+a^2+b^2\right)\right]-(a b+q)^2\right.}{2 \pi r_{+}\left[\left(r_{+}^2+a^2\right)\left(r_{+}^2+b^2\right)+a b q\right]}.
\end{align}
It is important to also find the chemical potentials associated to the angular momenta and the electric charge. The angular velocities $\Omega_{1}$ and $\Omega_{2}$ are given by
\begin{align} \label{eq:Omegageneral}
	\Omega_{1} & =\frac{a\left(r_{+}^2+b^2\right)\left(1+g^2 r_{+}^2\right)+b q}{\left(r_{+}^2+a^2\right)\left(r_{+}^2+b^2\right)+a b q}, \quad
	\Omega_{2} = \frac{b\left(r_{+}^2+a^2\right)\left(1+g^2 r_{+}^2\right)+a q}{\left(r_{+}^2+a^2\right)\left(r_{+}^2+b^2\right)+a b q}.
\end{align}
Via the Killing field that generates the horizon $\chi^{\mu} = \partial_{t} + \Omega_1 \partial_{\phi} + \Omega_2 \partial_{\psi}$, the electrostatic potential is
\begin{align} \label{eq:Phigeneral}
	\Phi &= \left. \chi^{\mu} A_{\mu} \right|_{r\to \infty} - \left.\chi^{\mu} A_{\mu} \right|_{r\to r_+} = \frac{\sqrt{3} g q r_{+}^2}{\left(r_{+}^2+a^2\right)\left(r_{+}^2+b^2\right)+a b q}.
\end{align}
Finally, the entropy can be computed via the area of the horizon
\begin{equation} \label{eq:CCLP-S}
	S = \frac{\pi^2\left[\left(r_{+}^2+a^2\right)\left(r_{+}^2+b^2\right)+a b q\right]}{2 G_{5}\Xi_a \Xi_b r_{+}}.
\end{equation}
We are particularly interested in extremal solutions, \textit{i.e.}, those where the inner and outer horizons coincide and the Hawking temperature vanishes. This condition is satisfied for
\begin{align} \label{eq:extcond}
	\Delta_r(r_+) = 0, \quad \partial_r \Delta_r(r_+) = 0.
\end{align}
For extremal solutions, the parametrization chosen in \cite{Chong:2005hr} with $\{a,b,m,q\}$ is not optimal as $\Delta_r$ is an order 6 polynomial and in order to write the thermodynamic quantities in a compact way, one would also require to use the parameter $r_+$ which is dependent on the others. We remedy this difficulty by finding a novel parametrization that automatically solves \eqref{eq:extcond}. The result is that the metric and thermodynamic quantities can be written compactly in terms of three parameters. We discuss this in detail in Section~\ref{sec:2der}.

Finally, the CCLP solution becomes supersymmetric when the constraint
\begin{equation}\label{susyCCLP}
q=\frac{m}{1+(a+b)g}
\end{equation}
is satisfied.
In general, the Lorentzian supersymmetric solutions are pathological unless one imposes an additional condition, corresponding precisely to the extremality condition $\partial_r \Delta_r(r_+) = 0$. For this supersymmetric and extremal solution the entropy can be expressed explicitly as a function of the charges as
\begin{equation}\label{eq:Ssusy0}
S=\pi\sqrt{4g^{-2}Q^2-\frac{\pi}{G_5 g^3}(J_{1}+J_{2})}\,.
\end{equation}
In addition, since the extremal supersymmetric solution only depends on two parameters, the charges satisfy a nonlinear constraint
\begin{equation}\label{eq:constr0}
	\left[2\sqrt{3}g^{-1}Q+\frac{\pi}{2G_{5}g^3}\right]\left[4g^{-2}Q^2-\frac{\pi}{G_{5}g^3}(J_1+J_2)\right]-\left(\frac{2Q}{\sqrt{3}g}\right)^3-\frac{2\pi}{G_{5}g^3}J_1J_2=0\, .
\end{equation}
These results can be matched with the field theory prediction for the entropy associated to the superconformal index as first shown in \cite{Cabo-Bizet:2018ehj, Benini:2018ywd, Choi:2018hmj}.  References \cite{Bobev:2022bjm, Cassani:2022lrk} found the higher-derivative corrections to these expressions and matched them again with the field theory results at subleading order in $(\mathfrak{a}-\mathfrak{c})$.  Here we intend to re-derive those results by analyzing the higher-derivative corrections to near-horizon geometries.

\section{Near-horizon geometries with $\mathrm{SL}(2,\mathbb{R})\times \mathrm{U}(1)\times \mathrm{U}(1)$ symmetry}
\label{sec:NHG}

\subsection{Ansatz}

The near-horizon geometries in which we are interested consist of an AdS$_2$ geometry fibered over two circles, with a total symmetry group $\mathrm{SL}(2,\mathbb{R})\times \mathrm{U}(1)\times \mathrm{U}(1)$. Near-horizon geometries of the same type have been studied in previous literature, \textit{e.g.} \cite{Kunduri:2007qy,Kunduri:2008rs}. However, here we perform an independent analysis that we deem to be more suitable for the applications we pursue. In particular, it is crucial to us not only analyzing the local form of the solution but also its global structure.

Without loss of generality, a metric and vector field with $\mathrm{SL}(2,\mathbb{R})\times \mathrm{U}(1)\times \mathrm{U}(1)$ symmetry can always be written as
\begin{align}\label{eq:metricansatz1}
ds^2&=W(x)\left(-r^2dt^2+\frac{dr^2}{r^2}\right)+\frac{W(x)}{G(x)}dx^2+\sum_{a,b=1}^2H_{ab}(x)\left(d\phi_a-\omega_a r dt\right)\left(d\phi_b-\omega_b r dt\right)\, ,\\
A&=\psi r dt+\sum_{a=1}^{2}A_a(x)\left(d\phi_a-\omega_a r dt\right)\, ,
\end{align}
where $a, b=1,2$, the coordinate $x$  is compact and $\phi_{1}$, $\phi_{2}$ are angular coordinates with periodicity $2\pi$.\footnote{As we will see, the metric functions $G(x)$ and $H_{ab}(x)$ have to satisfy certain properties in order to ensure absence of conical defects.} The quantities $\omega_{a}$ are constant --- required by the $\mathrm{SL}(2,\mathbb{R})$ symmetry --- and they are interpreted as chemical potentials conjugate to the angular momenta in each of the directions $\phi_1$ and $\phi_2$. They are related to the angular velocity of the horizon, but in the near-horizon geometry they lose that meaning as we lack an asymptotic region from where we can measure those velocities. 
Analogously, the constant $\psi$ is a variable conjugate to the electric charge and hence it is related to the electrostatic potential, although again the lack of an asymptotic region forbids us to interpret it as the usual electrostatic potential at infinity. Importantly, the three potentials $(\psi,\omega_1,\omega_2)$ are univocally determined from the near-horizon geometry once we impose regularity of the metric (absence of conical defects)  and of the gauge field (absence of divergences). We detail this below. 

The metric \req{eq:metricansatz1} still has some gauge freedom associated to reparametrizations of the $x$ coordinate. This allows us to set an extra condition in order to fix the gauge, and we find it useful to set
\begin{equation}\label{eq:metricdetgauge}
\frac{W(x)}{G(x)}\det\left[H_{ab}(x)\right]=\text{const.}
\end{equation}

Now, even though \req{eq:metricansatz1} is the most intuitive way of writing a metric with the given symmetries, it is not the most convenient form when solving the equations of motion. In fact, by performing a change of coordinates
\begin{equation}\label{eq:yphirel}
\begin{aligned}
y_1&=s_{1} \phi_1+s_{2}\phi_2\, ,\\
y_2&=s_{3} \phi_1+s_{4}\phi_2\, ,
\end{aligned}
\end{equation}
where the $s_i$ are constants, we can rearrange the solution as
\begin{equation}
\begin{aligned}\label{eq:metricansatz2}
ds^2&=W(x)\left(-r^2dt^2+\frac{dr^2}{r^2}\right)+\frac{W(x)}{G(x)}dx^2+\frac{G(x)}{W(x)B(x)}dy_1^2+B(x)\left(dy_2+J(x) dy_1-r dt\right)^2\, ,\\
A&=\psi r dt+\left(\hat{A}_1(x)-\psi J(x)+k\right)dy_1+\left(\hat{A}_2(x)+\psi \right)\left(dy_2+J(x) dy_1- r dt\right)\, ,
\end{aligned}
\end{equation}
where we have set $g_{xx}\left(g_{y_1y_1}g_{y_2y_2}-(g_{y_1y_2})^2\right)=1$ following the gauge condition \req{eq:metricdetgauge}. Furthermore, we have decomposed the $A_{y_1}$ and $A_{y_2}$ components of the gauge field in a way in which $\psi$ as well as the new constant $k$ are gauge parameters. Thus, they do not appear in the equations of motion, but nevertheless they are fixed by regularity, as we show next.

\subsection{Angular coordinates and chemical potentials  }
In practice, once a solution of the form \req{eq:metricansatz2} has been found, one must then undo the change of variables \req{eq:yphirel} in order to identify the angular coordinates $\phi_{a}$ and the chemical potentials $\omega_{a}$. However, one does not know a priori this change of variables and one must derive it entirely from the properties of the near-horizon metric \req{eq:metricansatz2}.  Here we show how to do this in full generality upon the assumption of $\mathbb{S}^{3}$ topology, which is the topology of the CCLP black hole horizon \cite{Chong:2005hr}. 

As was mentioned, the coordinate $x$ is compact and thus it has a range $x\in [x_1, x_2]$, which is determined by the vanishing of $G(x)$ at those points. However, by doing a linear change of variable $x\rightarrow \alpha+\beta x$ we can always set $x_{1}=-1$, $x_{2}=1$. Observe that the effect of this change of coordinates can be reabsorbed by performing $y_{1}\rightarrow y_{1}/\beta^2$ and redefining $\left\{G(x), J(x), \hat{A}_{1}(x), k\right\}\rightarrow \left\{\beta^2 G(x), \beta J(x), \beta \hat{A}_{1}(x), \beta k\right\}$. Hence, without loss of generality we assume $x\in [-1, 1]$, which allows us to interpret $x=\cos\tilde{\theta}$ with $\tilde{\theta}\in [0,\pi]$, although we will work in terms of $x$ for convenience.

The points $x=\pm 1$ are then identified as the poles of the horizon. Near those points the angular coordinates are identified by demanding regularity of the metric, \textit{i.e.}, absence of conical defects. Let us examine the behavior of the metric near those points for a constant $t$ and $r$ slice
\begin{align}\label{eq:metric3}
ds^2_{3}&=\frac{W(x)}{G(x)}dx^2+\frac{G(x)}{W(x)B(x)}dy_1^2+B(x)\left(dy_2+J(x) dy_1\right)^2\, .
\end{align}
First, we introduce the coordinate $z$ such that

\begin{equation}
dz=dx\sqrt{\frac{W(x)}{G(x)}}\, .
\end{equation}
Integrating this relation near $x=\pm 1$ yields
\begin{equation}
x=1+\frac{z_{+}^2 G'(1)}{4 W(1)}+\mathcal{O}(z_{+}^4)\, ,\quad x=-1+\frac{z_{-}^2 G'(-1)}{4 W(-1)}+\mathcal{O}(z_{-}^4)\, .
\end{equation}
We then rewrite the $y_{a}$ coordinates in terms of the angular coordinates $\phi_{a}$ as in \req{eq:yphirel}. We demand the periodicity of $\phi_{a}$ to be $2\pi$ and $s_i$ are undetermined coefficients. We series expand the metric around $x=-1$ and $x=+1$ up to order $z^2$. Then, in order for the geometry to be regular, the metric around those points must take the form
\begin{align}
ds^2_{3}(x\rightarrow +1)&=dz_{+}^2+z_{+}^2\left(d\phi_{2}+C_{+} d\phi_{1}\right)^2+(D_{+}+E_{+}z_{+}^2) d\phi_{1}^2+\mathcal{O}(z_{+}^4)\, ,\\
ds^2_{3}(x\rightarrow -1)&=dz_{-}^2+z_{-}^2\left(d\phi_{1}+C_{-}d\phi_{2}\right)^2+(D_{-}+E_{-}z_{-}^2)  d\phi_{2}^2+\mathcal{O}(z_{-}^4)\, ,
\end{align}
where $C_{\pm}$, $D_{\pm}$, $E_{\pm}$ are constants.
In particular, this means that
\begin{equation}
g_{\phi_2\phi_2}=z_{+}^2+\mathcal{O}(z_{+}^4)\, ,\quad g_{\phi_1\phi_2}=0+\mathcal{O}(z_{+}^2)
\end{equation}
near $x=1$ while 
\begin{equation}
g_{\phi_1\phi_1}=z_{-}^2+\mathcal{O}(z_{-}^2)\, ,\quad g_{\phi_1\phi_2}=0+\mathcal{O}(z_{-}^2)
\end{equation}
near $x=-1$. These are four conditions that fix the four coefficients $s_i$. We find the unique solution up to orientation changes to be
\begin{equation}\label{eq:angles}
	\begin{aligned}
		y_{1}&=2\mathcal{H}(1)\phi_{2}+2\mathcal{H}(-1)\phi_{1}\, ,\\
		y_{2}&=-2J(1)\mathcal{H}(1)\phi_{2}-2J(-1)\mathcal{H}(-1)\phi_1\, ,
	\end{aligned}
\end{equation}
where we have defined
\begin{align}
	\mathcal{H}(x) = \frac{\sqrt{B(x)}W(x)}{G'(x)}\, .
\end{align}
From the relations \eqref{eq:angles} we find the generators of rotations
\begin{equation}\label{eq:killinggen}
	\begin{aligned}
		\partial_{\phi_{1}}&=2\mathcal{H}(-1)\left(\partial_{y_1}-J(-1)\partial_{y_{2}}\right)\, ,\\
		\partial_{\phi_{2}}&=2\mathcal{H}(1)\left(\partial_{y_1}-J(1)\partial_{y_{2}}\right)\, ,
	\end{aligned}
\end{equation}
which we need in order to compute the angular momenta. We can also obtain the area element of the horizon in terms of the angular coordinates,
\begin{equation}
	d\Sigma_{3}=dx\wedge dy_{1}\wedge dy_{2}=4dx\wedge d\phi_{1}\wedge d\phi_{2} \mathcal{H}(1)\mathcal{H}(-1)(J(-1)-J(1))\, ,
\end{equation}
which allows us to compute the area right away
\begin{equation}\label{eq:areagen}
	\mathcal{A}=32\pi^2 |\mathcal{H}(1)\mathcal{H}(-1)(J(1)-J(-1))|\,.
\end{equation}
After identifying the angular coordinates, we can apply the change of coordinates \req{eq:angles} in \req{eq:metricansatz2} and rewrite the metric in the canonical form \req{eq:metricansatz1}. This allows us to read off the chemical potentials $\omega_{a}$
\begin{align}\label{eq:omegagen}
	\omega_{1}=\frac{1}{2\mathcal{H}(-1) (J(1) -J(-1))}\, ,\quad \omega_{2}=-\frac{1}{2\mathcal{H}(1)(J(1)-J(-1))}\, .
\end{align}
On the other hand, the electrostatic potential is determined by the regularity of the gauge field at the poles of the horizon. This can be analyzed by using \req{eq:metricansatz2}. Since the metric component $g_{y_1 y_1}$ vanishes at $x=\pm 1$, the component of the gauge field $A_{y_{1}}$ must also vanish at those points --- otherwise the gauge field would be singular, as one can check by computing its norm. Thus, we impose
\begin{equation}
\hat{A}_1(x)-\psi J(x)+k\Big|_{x=\pm1}=0
\end{equation}
and we find

\begin{align}\label{eq:psigen}
\psi=&\frac{\hat{A}_{1}(1)-\hat{A}_{1}(-1)}{J(1)-J(-1)}\, ,\\
k=&\frac{J(-1) \hat{A}_{1}(1)-J(1) \hat{A}_{1}(-1)}{J(1)-J(-1)}\, .
\end{align}
In this way, all the chemical potentials are intrinsically determined by the regularity of the solution.

\section{Near-horizon geometry of the two-derivative theory}
\label{sec:2der}
\subsection{Solving the equations of motion  }
The near-horizon extremal geometry in minimal gauged supergravity can be obtained by taking the appropriate near-horizon limit of the CCLP solution. However, the resulting metric and gauge potential are considerably involved and not in the form of \req{eq:metricansatz2}. Another obstacle is that in the case of higher-derivative gravity, we do not know the complete solution and we must work directly in the near-horizon geometry. Thus, here we solve directly the equations of motion of the two-derivative theory for the ansatz \req{eq:metricansatz2}. This allows us to extract some insight on how the equations may be solved and then to implement it in the case when higher derivatives are turned on. Along the way, we use inspiration from the CCLP metric in some steps of the derivation.

First, it is useful to express the equations in the frame
\begin{equation}\label{framedef}
\begin{aligned}
e^{0}&=\sqrt{W}r dt\, ,\quad & e^{1}&=\sqrt{W}\frac{dr}{r}\,,\quad e^{2}=\sqrt{\frac{W}{G}}dx\, ,\\
\quad e^{3}&=\sqrt{\frac{G}{W B}}dy_{1}\, ,\quad & e^{4}&=\sqrt{B}\left(dy_2+J dy_1-r dt\right)\,.
\end{aligned}
\end{equation}
We note that the functions $J(x)$ and $\hat{A}_{1}(x)$ always appear with at least one derivative acting upon them, and we can instead consider our variables to be $J'(x)$ and $\hat{A}_{1}'(x)$, reducing the order of the equations. In fact, it proves interesting to perform the change of variables
\begin{equation}\label{eq:JAprime}
J'=\frac{j}{WB}\, ,\quad \hat{A}_{1}'=\frac{\zeta B-j\hat{A}_{2}}{WB^2}\, ,
\end{equation}
where we have introduced the functions $j(x)$ and $\zeta(x)$. 
In terms of these variables, the $\mathcal{E}_{3}$ component of the Maxwell equation and the $\mathcal{E}_{34}$ component of the Einstein equation yield, respectively
\begin{align}\label{eq:zetajeq}
3 \zeta'-2 \sqrt{3} \hat{A}_{2} \hat{A}_{2}'=0\, ,\quad
j'+\zeta \hat{A}_{2}'=0\, .
\end{align}
These can be integrated immediately to find 
\begin{equation}
\zeta=\kappa_{1}+\frac{1}{\sqrt{3}}\hat{A}_{2}^2\, ,\quad j=\kappa_{2}-\kappa_{1}\hat{A}_{2}-\frac{1}{3\sqrt{3}}\hat{A}_{2}^2\, ,
\end{equation}
where $\kappa_{1}$ and $\kappa_{2}$ are integration constants.  On the other hand, we have the following combination of components of Einstein equations
\begin{equation}\label{eq:Weq}
\mathcal{E}_{00}-\mathcal{E}_{22}=-\frac{G''-24 g^2 W+2}{2 W}=0\, ,
\end{equation}
so that we can express $W$ in terms of $G$ as
\begin{equation}\label{eq:Wsol}
W=\frac{G''+2}{24 g^2}\, .
\end{equation}
Another interesting combination of Einstein equations is the following
\begin{equation}\label{eq:Beq}
\begin{aligned}
&\mathcal{E}_{00}+\mathcal{E}_{22}+\mathcal{E}_{33}+\mathcal{E}_{44}=-\frac{3}{2BW^2}\Bigg[B^2+\frac{B}{3 W} \left(W \left(2 \hat{A}_2^2-3 \left(G' W'+G W''\right)\right)\right.\\
&\left.+3 G W'{}^2+24 g^2 W^3-6 W^2\right)+\frac{1}{9} \left(3 G W \hat{A}_2'{}^2+\hat{A}_2^4+2 \sqrt{3} \hat{A}_2^2 \kappa_1+3 \kappa_1{}^2\right)\Bigg]=0\, .
\end{aligned}
\end{equation}
This equation determines algebraically $B$ in terms of $G$ and $ \hat{A}_2$, since $W$ is also determined by $G$. The remaining Einstein equations and the $\mathcal{E}_{4}$ component of Maxwell equations form a highly nonlinear system of equations of $G$ and $\hat{A}_{2}$. Furthermore, these contain up to six derivatives of $G$, making finding the solution highly challenging. To remedy the difficulty, we take inspiration from the near-horizon limit of the CCLP metric and consider the following ansatz for these functions
\begin{equation}
G=(1-x^2)(\kappa_{3}+\kappa_{4}x)\, ,\quad \hat{A}_{2}=\frac{\kappa_{5}}{W}\, ,
\end{equation}
with undetermined constants $\kappa_{i}$. With this, we can obtain an explicit expression for $B$ from \req{eq:Beq}. For general values of $\kappa_{i}$, the solution for $B$ contains square roots --- being the solution of the quadratic equation \eqref{eq:Beq}. However, we constraint the $\kappa_{i}$ constants in order to form a perfect square inside the square root, so that $B$ becomes a rational function of $x$. It turns out that doing so already solves most of the components of the equations of motion. The remaining components simply set an additional constraint on the remaining integration constants $\kappa_{i}$.

Finally, $J$ and $\hat{A}_{1}$ can be obtained by integrating \req{eq:JAprime}. 
Thus, the solution is determined up to solving the constraint equations between the $\kappa_{i}$ coefficients. We do not explicitly show the constraint here due to its complicated nature. In fact, one of the most challenging aspects about the solution is finding an appropriate parametrization in which these coefficients take a simple form. We have been able to identify a set of three parameters $(X,Y, Z)$ that allow us to express the solution in an explicit and fairly compact form. The result reads

\begin{equation}\label{eq:solution2der}
\begin{aligned}
W=&\frac{\Delta }{4 g^2 \left(3-R^2\right)}\, ,\\
G=&\left(1-x^2\right)\frac{ (x-1) X^2-(x+1) Y^2+2
   Z^2}{3-R^2}\, ,\\
   B=&\frac{\Delta }{4 g^2
   \left(3-R^2\right) }-\frac{\left(1-R^2\right) }{2 g^2
   \left(3-R^2\right)^2 \Delta ^2}\bigg[\left(1-R^2-2 X Y Z\right)^2\\
   &+\frac{\Delta}{2}  (-1+X-Y-Z) (1+X+Y-Z) (1+X-Y+Z) (-1+X+Y+Z)\bigg]\, ,\\
   J=&-\frac{4\sqrt{2} g \sqrt{1- R^2}}{\left(3-R^2\right) \left(X^2-Y^2\right)
 B  \Delta ^2}\bigg[\left(1-R^2-2 X Y Z\right)^2\\
 &-\Delta\left(1-R^2+2Z^2(X^2+Y^2)+2X^2Y^2+X Y Z
   \left(1+R^2\right)\right) \bigg]\, ,\\
   \hat{A}_{1}=&\frac{2 \sqrt{3} \left(1-R^2-2 X Y Z\right) \left(\left(1-R^2\right)
   \left(1-R^2-2 X Y Z\right)-\Delta ^2\right)}{\left(3-R^2\right)
    (X^2-Y^2) B \Delta ^2}\, ,\\
\hat{A}_{2}=&-\frac{\sqrt{\frac{3}{2}} \left(1-R^2-2 X Y Z\right)
   \sqrt{1-R^2}}{g \left(3-R^2\right) \Delta }\, ,
\end{aligned}
\end{equation}
where we have introduced the notation\footnote{This $R$ should not be confused with the Ricci scalar. }

\begin{align}
\Delta&=1-x \left(X^2-Y^2\right)-Z^2\, ,\\
R&=\sqrt{X^2+Y^2+Z^2}\, .
\end{align}
We note that the expressions for $J$ and $\hat{A}_{1}$ are divergent for $Y\rightarrow X$, but the divergent piece is a constant term that can be reabsorbed in the choice of integration constant from \req{eq:JAprime}, so the solution is actually regular at $X=Y$.\footnote{We discuss the case of $X=Y$ in Subsection~\ref{sec:equalJsol} and the solution of the various functions in the ansatz is written in \eqref{eq:solution2derequalJ}.}

From these expressions it is quite clear that a real Lorentzian solution only exists if $R\le 1$ and therefore in terms of the $(X,Y,Z)$ parameters the space of solutions lies within the unit ball in $\mathbb{R}^{3}$. Note that at $R=1$ the horizon area is zero yielding a naked singularity. For this reason, we choose to consider only the regime $R < 1$.
These parameters turn out to be related to the original parameters $(a, b, q, m, r_{+})$ of the CCLP solution \cite{Chong:2005hr} via
\begin{equation}\label{CCLPrel}
	\begin{aligned}
		a=&\frac{X}{g Z}\, , & \quad m=&\frac{3-2R^2-R^4+4 X^2 Y^2+4 X^2Z^2+4 Y^2 Z^2}{8g^2 Z^4}\, , \quad &
		r_{+}^2=&\frac{1-R^2}{2g^2 Z^2}\, ,\\
		b=&\frac{Y}{g Z}\, , &
		q=&\frac{1-R^2-2 X Y Z}{2 g^2 Z^3}\, .
	\end{aligned}
\end{equation}
This can be checked by comparing some of the metric components at the horizon taking into account that $x=\cos(2\theta)$\footnote{In particular, the $g_{xx}$ and $g_{rr}$ components are the easiest ones to check.} and by comparing the expressions for the charges and entropy that we provide below.  In addition, we verify that the relations \req{CCLPrel} automatically solve the equation that determines the position of the horizon as well as extremality condition for CCLP black holes \req{eq:extcond}. Therefore, \eqref{CCLPrel} provides an interesting and useful parametrization of the space of extremal CCLP solutions.

\begin{figure}
	\centering
\includegraphics[scale=0.8]{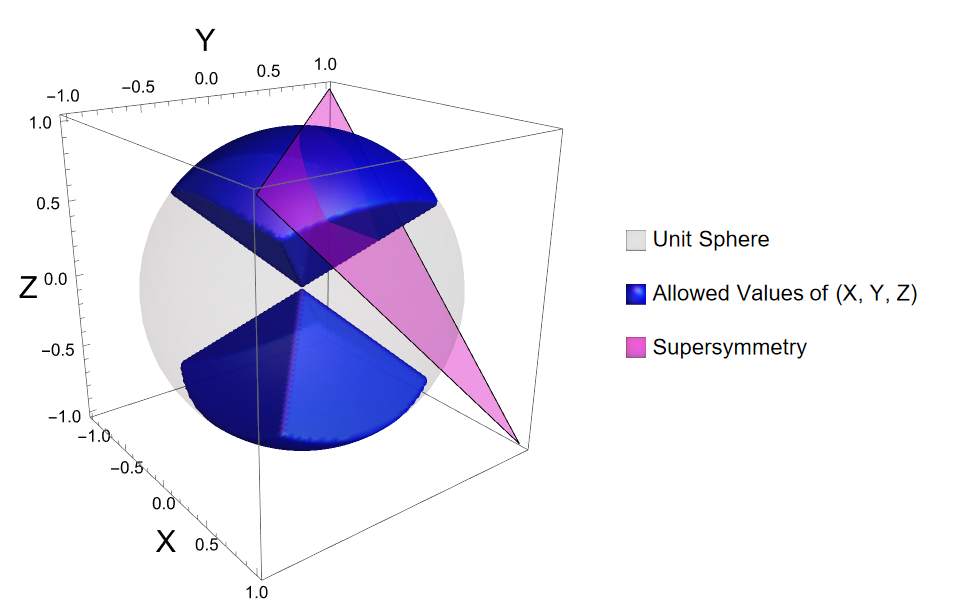}
\caption{The allowed region in the parameter space $(X,Y,Z)$ of extremal black holes is given by two sections of a unit sphere as shaded in blue. The unit sphere is denoted in gray. Supersymmetry is allowed on the pink plane defined by $X+Y+Z=1$. Extremal and supersymmetric black holes are allowed for regions where the supersymmetric plane intersects with the extremal solutions. Note that the supersymmetric plane only intersects in one of the two blue regions.}
\label{fig:spaceofsolutions}
\end{figure}

Besides the constraint $R<1$, these relations also inform us that we should restrict our space of solutions to $X^2<Z^2$ and $Y^2<Z^2$, which are equivalent to the usual constraints $(ag)^2<1$, $(bg)^2<1$. In this way, the space of solutions is restricted to two pyramidal spherical caps as shown in Figure~\ref{fig:spaceofsolutions}.
Finally, in terms of the $(X,Y, Z)$ parameters the supersymmetric condition \req{susyCCLP} takes the simple form
\begin{equation}\label{susyXYZ}
\quad X+Y+Z=1\, .
\end{equation}
In this case, we can write these variables explicitly in terms of $a$ and $b$ as
\begin{equation}\label{susyXYZ2}
X^{\star}=\frac{ag}{1+(a+b)g}\, ,\quad Y^{\star}=\frac{bg}{1+(a+b)g}\, ,\quad Z^{\star}=\frac{1}{1+(a+b)g}\, ,
\end{equation}
where the star $\star$ is meant to represent supersymmetric extremal values. 
\subsection{Thermodynamics}
With the solution \req{eq:solution2der}, we evaluate the various thermodynamic quantities of the black hole. Using the general formulas \req{eq:omegagen} and \req{eq:psigen}, it is straightforward to obtain the near-horizon chemical potentials
\begin{equation}\label{potentials2d}
\begin{aligned}
\omega_{1}=&-\frac{\sqrt{2} \left(Z^2-X^2\right) \left[X Y^2 Z+\left(2 Y+XZ\right)  \left(X^2+Z^2-1\right)\right]}{\sqrt{1-R^2}
   \left(3-R^2\right) \left[2 X Y+Z\left(1+X^2 +Y^2 -Z^2\right)\right]}\, ,\\
\omega_{2}=&-\frac{\sqrt{2} \left(Z^2-Y^2\right) \left[Y X^2 Z+\left(2 X+YZ\right)  \left(Y^2+Z^2-1\right)\right]}{\sqrt{1-R^2}
   \left(3-R^2\right) \left[2 X Y+Z\left(1+X^2 +Y^2 -Z^2\right)\right]}\, ,\\
\psi=&\frac{\sqrt{\frac{3}{2}} \left(1-R^2-2 X Y Z\right) \left[Z \left(1-R^2\right)-2 X Y\right]}{g \sqrt{1-R^2}
   \left(3-R^2\right)\left[2 X Y+Z\left(1+X^2 +Y^2 -Z^2\right)\right]}\, .
\end{aligned}
\end{equation}
We can also obtain the black hole entropy $S=\mathcal{A}/(4G_5)$ by direct application of \req{eq:areagen}, yielding 
\begin{equation}\label{entropy2d}
S=\frac{\pi ^2 \sqrt{1-R^2} \left(Z \left(X^2+Y^2+1\right)+2 X Y-Z^3\right)}{4 \sqrt{2} G_5 g^3 (Z^2-X^2)(Z^2-Y^2)}\, .
\end{equation}
Then, the charge and angular momenta can be obtained by evaluating the integrals \eqref{eq:CCLP-Q}, \eqref{eq:CCLP-Ja} and \eqref{eq:CCLP-Jb}, respectively, where in the latter we have to use the Killing vectors defined in \req{eq:killinggen}. This yields
\begin{equation}\label{charges2d}
\begin{aligned}
Q=&\frac{\sqrt{3} \pi  Z \left(1-R^2-2 X Y Z\right)}{8 G_5 g^2 \left(Z^2-X^2\right)\left(Z^2-Y^2\right)}\, , \\
J_{1}=&-\frac{\pi}{16 G_5 g^3 \left(Z^2-X^2\right)^2 (Z^2-Y^2)}  \Big[2 X^3 Z \left(Y^2-Z^2+1\right)+2 X^2 Y \left(Y^2+2 Z^2-1\right)\\
&+2 X^4 Y+X^5 Z+X Z \left(Y^2+Z^2-1\right) \left(Y^2+Z^2+3\right)+2 Y Z^2 \left(Y^2+Z^2-1\right)\Big]\,  ,\\
J_{2}=&J_1(X\leftrightarrow Y)\,.
\end{aligned}
\end{equation}
These results deserve several comments. First, the expressions for the charges and entropy coincide with those of \cite{Chong:2005hr}, specified for extremal solutions, upon using the relations \req{CCLPrel}. Obviously this includes the supersymmetric case given by \req{susyXYZ2}. Second, the near-horizon potentials and the charges satisfy a \textit{near-horizon version} of the first law of thermodynamics \cite{Hajian:2013lna}  that reads 
\begin{equation}\label{1stlawhorizon}
\frac{1}{2\pi}dS=\omega_{1}dJ_{1}+\omega_{2}dJ_{2}+\psi dQ\, .
\end{equation}
This allows us to understand the meaning of the potentials in \eqref{potentials2d}. In general, the complete first law of thermodynamics for the free energy $F$ in the canonical ensemble reads
\begin{equation}
dF=-S dT+\sum_{i=1}^{2}\Omega_{i}dJ_{i}+\Psi dQ\, ,
\end{equation}
where $\Omega_{i}$ and $\Psi$ are the angular velocities and the electrostatic potential, respectively, which are given by the difference between values at the near-horizon and infinity. From this, taking the second derivatives of the free energy and using that they commute, we can derive the relations
\begin{equation}
\left(\frac{\partial S}{\partial J_{i}}\right)_{T, \,Q}=-\left(\frac{\partial \Omega_{i}}{\partial T}\right)_{J_{i}\, ,Q}\, ,\quad \left(\frac{\partial S}{\partial Q}\right)_{T, \,J_{i}}=-\left(\frac{\partial \Psi}{\partial T}\right)_{J_{i}\, ,Q}\, ,
\end{equation}
where the subindices denote which variables are kept fixed when taking the derivative. On the other hand, \req{1stlawhorizon} is equivalent to
\begin{equation}
\left(\frac{\partial S}{\partial J_{i}}\right)_{T=0, \,Q}=2\pi \omega_{i}\, ,\quad \left(\frac{\partial S}{\partial Q}\right)_{T=0, \,J_{i}}=2\pi \psi\, ,
\end{equation}
Comparing both expressions allows us to identify the near-horizon potentials as ``fugacities''
\begin{equation}
\omega_{i}=-\frac{1}{2\pi} \frac{\partial \Omega_{i}}{\partial T}\bigg|_{T=0}\, ,\quad \psi=-\frac{1}{2\pi} \frac{\partial \Psi}{\partial T}\bigg|_{T=0}\, ,
\end{equation}
where $\Omega_{i}$ and $\Psi$ have to be understood as variables of $T$, $J_i$ and $Q$. We can rewrite these relations in a suggestive way as 
\begin{equation} \label{eq:potdef2}
\omega_{i}=-\lim_{T\rightarrow 0} \frac{\Omega_{i}-\Omega_{i}^{\rm ext}}{2\pi T}\, ,\quad \psi=- \lim_{T\rightarrow 0} \frac{\Psi-\Psi^{\rm ext}}{2\pi T}\, ,
\end{equation}
where $\Omega_{i}^{\rm ext}$ and $\Psi^{\rm ext}$ represent the values at extremality. 
This is very similar to the definition of the supersymmetric potentials which are often used in the literature as in \cite{Silva:2006xv,Dias:2007dj,Sen:2008yk, Cabo-Bizet:2018ehj, Choi:2018hmj}, with the difference that, in those cases, the whole expression is evaluated on the (complex) supersymmetric solution. In fact, in the supersymmetric case, our potentials \req{potentials2d} satisfy the constraint
\begin{equation} \label{eq:susyconstr}
\omega_{1}+\omega_{2}-\sqrt{3}g\psi\Big|_{\rm susy}=0\,.                                                                                                                                                    
\end{equation}
This is entirely analogous to the complex constraint satisfied by the supersymmetric potentials \cite{Cabo-Bizet:2018ehj, Benini:2018ywd, Choi:2018hmj}, although in that case the right-hand side of the expression would be $+i$ (in our conventions). Such difference appears due to the different order in which the supersymmetric and extremal limits are taken. In our case, we consider the extremal limit first and then impose supersymmetry, so that the solution is Lorentzian and the potentials are real. In the next sections, we will use \req{eq:susyconstr} as the defining property of a supersymmetric solution. 

\section{Higher-derivative corrections}
\label{sec:higher}
\subsection{The setup} \label{subsec:HDsetup}
We now study the higher-derivative corrections in the action \req{eq:action} to the solution \req{eq:solution2der}. Let us first explain the logic behind the resolution of the equations of motion. We work perturbatively at first order in $\alpha$, so that we assume that the metric and gauge field are expanded as
\begin{equation}
g_{\mu\nu}=g_{\mu\nu}^{(0)}+\alpha g_{\mu\nu}^{(1)}+\mathcal{O}(\alpha^2)\, ,\quad A_{\mu}=A_{\mu}^{(0)}+\alpha A_{\mu}^{(1)}+\mathcal{O}(\alpha^2)\, ,
\end{equation}
where $g_{\mu\nu}^{(0)}$ and $A_{\mu}^{(0)}$ satisfy the two-derivative equations of motion. 
This allows us to write the Einstein and Maxwell equations in \eqref{eq:eoms} as
\begin{align}
G_{\mu\nu}+6 g^2 g_{\mu\nu}-\frac{1}{2}F_{(\mu|\rho}F_{|\nu)}{}^{\rho}+\frac{1}{8}g_{\mu\nu}F^2&=\alpha T_{\mu\nu}^{\rm eff}(g^{(0)},A^{(0)})\, ,\\
-\nabla_{\mu}\mathcal{F}^{\mu\nu}+\frac{1}{4\sqrt{3}}\epsilon^{\nu\alpha\beta\gamma\delta}F_{\alpha\beta}F_{\gamma\delta}&=\alpha J^{\nu}_{\rm eff}(g^{(0)},A^{(0)})\, ,
\end{align}
where in the right-hand side we collect all the terms proportional to $\alpha$ and we evaluate them on the zeroth-order solution. Thus, the equations that we have to solve are equivalent to the original two-derivative equations with a fixed source. The process of solving the equations of motion can in principle be carried out along the same lines as in Section~\ref{sec:2der}. We consider the ansatz \req{eq:metricansatz2} and we introduce the functions $j(x)$ and $\zeta(x)$ as in \req{eq:JAprime}. Then, we expand all the functions linearly in $\alpha$ as 
\begin{equation}\label{funcexpan}
\begin{aligned}
W&=W^{(0)}+\alpha g^2 W^{(1)}\, ,\quad & G&=G^{(0)}+\alpha g^2 G^{(1)}\, ,\quad & B&=B^{(0)}+\alpha g^2 B^{(1)}\, ,\\
j&=j^{(0)}+\alpha g^2 j^{(1)}\, ,\quad & \zeta&=\zeta^{(0)}+\alpha g^2 \zeta^{(1)}\, ,\quad & \hat{A}_{2}&=\hat{A}_{2}^{(0)}+\alpha g^2 \hat{A}_{2}^{(1)}\, ,
\end{aligned}
\end{equation}
where the functions with the $(0)$ superscript are those in \req{eq:solution2der}.  We arrive once more to the equations \req{eq:zetajeq}, \req{eq:Weq}, \req{eq:Beq}, but in this case the right-hand side to these equations depends only on the zeroth-order solution. These four equations \req{eq:JAprime},  \req{eq:zetajeq}, \req{eq:Weq}, \req{eq:Beq} can be integrated completely to find $j^{(1)}$, $\zeta^{(1)}$, $W^{(1)}$ and $B^{(1)}$ in terms of $G^{(1)}$ and $\hat{A}_{2}^{(1)}$. The two remaining variables  $G^{(1)}$ and $\hat{A}_{2}^{(1)}$ satisfy a system of two coupled linear differential equations with up to sixth-order derivatives. Solving these equations is highly challenging, but it becomes much simpler in the case of equal angular momenta or when the difference between the angular momenta is small, as we consider in the next subsections. Finally, once the zeroth- and first-order solutions have been found for the variables in \req{funcexpan}, it is straight forward to obtain
\begin{equation}\label{funcexpan2}
J=J^{(0)}+\alpha g^2 J^{(1)}\, ,\quad \hat{A}_{1}=\hat{A}_{1}^{(0)}+\alpha g^2 \hat{A}_{1}^{(1)}\,,
\end{equation}
from \req{eq:JAprime}.

In order to specify a solution, we also need to fix boundary conditions. 
Indeed, in our analysis, we find some integration constants that correspond to a relabeling of the parameters of the two-derivative solution as $(X, Y, Z)\rightarrow (X+\alpha \delta X, Y+\alpha \delta Y, Z+\alpha \delta Z)$. Thus, in order to fix the solution we work in the grand canonical ensemble, so that we fix $\omega_{1}$, $\omega_{2}$ and $\psi$ to be given by the same expressions \req{potentials2d} of the two-derivative solution. We recall that these potentials are given by the general formulas \req{eq:omegagen} and \req{eq:psigen}, so when we expand the functions as in \req{funcexpan} and \req{funcexpan2}, the potentials take the form
\begin{equation}
\omega_{i}=\omega_{i}^{(0)}\left(1+\alpha g^2 \delta\omega_{i}\right)\, ,\quad \psi=\psi^{(0)}\left(1+\alpha g^2 \delta\psi\right)\, ,
\end{equation}
where
\begin{equation}\label{correctionpotentials}
\begin{aligned}
\delta\omega_{1}=&\frac{G^{(1)}{}'(-1)}{G^{(0)}{}'(-1)}-\frac{B^{(1)}(-1)}{2 B^{(0)}(-1)}-\frac{W^{(1)}(-1)}{W^{(0)}(-1)}-\frac{J^{(1)}(1)-J^{(1)}(-1)}{J^{(0)}(1)-J^{(0)}(-1)}\, ,\\
\delta\omega_{2}=&\frac{G^{(1)}{}'(1)}{G^{(0)}{}'(1)}-\frac{B^{(1)}(1)}{2 B^{(0)}(1)}-\frac{W^{(1)}(1)}{W^{(0)}(1)}-\frac{J^{(1)}(1)-J^{(1)}(-1)}{J^{(0)}(1)-J^{(0)}(-1)}\, ,\\
\delta\psi=&\frac{\hat{A}_{1}^{(1)}(1)-\hat{A}_{1}^{(1)}(-1)}{\hat{A}_{1}^{(0)}(1)-\hat{A}_{1}^{(0)}(-1)}-\frac{J^{(1)}(1)-J^{(1)}(-1)}{J^{(0)}(1)-J^{(0)}(-1)}\,.
\end{aligned}
\end{equation}
Thus, we impose the conditions
\begin{equation}\label{bdycond}
\delta\omega_{1}=\delta\omega_{2}=\delta\psi=0\, ,
\end{equation}
which in turn imply boundary conditions on the functions of the solution at $x=\pm 1$.

\subsection{Equal angular momenta  }\label{sec:equalJsol}
When the black hole has equal angular momenta, corresponding to $X=Y$ in the parametrization introduced in \req{eq:solution2der}, the $\mathrm{U}(1)\times \mathrm{U}(1)$ symmetry of the solution is enhanced to $\mathrm{SU}(2)\times \mathrm{U}(1)$. As a result, all the functions in the ansatz have a fixed functional dependence. We have $G(x)\propto (1-x^2)$, while the rest of functions in \req{funcexpan} must be constant. In turn, the functions $J(x)$ and $\hat{A}_{1}(x)$ become linear in $x$ up to an irrelevant additive constant, as only derivatives of $J(x)$ and $\hat{A}_{1}(x)$ appear in the equations. This allows us to express the solution as 
\begin{equation}
\begin{aligned}\label{eq:solution2derequalJ}
W=&\frac{1-Z^2}{4 g^2 \left(3-2 X^2-Z^2\right)}\left(1+\alpha g^2 \delta W\right)\, ,\\
G=&\frac{2 \left(1-x^2\right) \left(Z^2-X^2\right)}{3-2 X^2-Z^2}\left(1+\alpha g^2 \delta G\right)\, ,\\
   B=&\frac{X^2 \left(2-2 X^2-Z^2-Z\right)^2}{g^2 (1-Z)^2 \left(3-2 X^2-Z^2\right)^2}\left(1+\alpha g^2 \delta B\right)\, ,\\
   J=&x\frac{2 \sqrt{2} g^3 \sqrt{1-2 X^2-Z^2} \left(2 X^2+Z-Z^2\right)
   \left(3-2 X^2-Z^2\right)^2}{X^2 (1+Z) \left(2-2 X^2-Z^2-Z\right)^2}\left(1+\alpha g^2 \delta J \right)\, ,\\
   \hat{A}_{1}=&-x\frac{2 \sqrt{3} g^2 \left(1-2 X^2-Z\right) \left(2 X^2+(-1+Z)
   Z\right) \left(3-2 X^2-Z^2\right)}{X^2 \left(2-2 X^2-Z^2-Z\right)^2}\left(1+\alpha g^2 \delta\hat{A}_{1} \right) ,\\
\hat{A}_{2}=&-\frac{\sqrt{\frac{3}{2}} \left(1-2 X^2-Z\right) \sqrt{1-2 X^2-Z^2}}{g
   (1-Z) \left(3-2 X^2-Z^2\right)}\left(1+\alpha g^2 \delta \hat{A}_{2}\right)\, ,
\end{aligned}
\end{equation}
where the leading terms correspond to the two-derivative solution \req{eq:solution2der} with $X=Y$ (in the case of $J$ and $\hat{A}_{1}$ we took the $Y\rightarrow X$ limit and removed the constant term) and the higher-derivative corrections are captured by the constant coefficients $\delta W$, $\delta G$, $\delta B$, $\delta J$, $\delta \hat{A}_{i}$. Thus, the higher-derivative Einstein and Maxwell equations become a linear system of algebraic equations for these coefficients whose solution is straightforward to obtain. Moreover, we also use the equations in \req{bdycond}, which impose additional constraints on the coefficients. When we do so, the solution is completely fixed. The full expressions for these coefficients are somewhat lengthy and thus we provide them in Appendix~\ref{app:solution}.

\subsubsection*{Zero chemical potential}
There are two interesting limits of this solution. One is the supersymmetric limit, which is contained in the analysis of the next section. The other is the case of zero chemical potential. By using \req{potentials2d} with $Y=X$, we can see that imposing $\psi=0$ amounts to setting 
\begin{equation}
Z=1-2X^2\, .
\end{equation}
Naturally, in the two-derivative solution, this also implies the vanishing of the gauge field, $A_{\mu}=0$, so that the solution is purely gravitational. Interestingly, this is no longer the case when the higher-derivative corrections are included. Indeed, at first order in $\alpha$, the solution with $\psi=0$ and equal angular momenta reads
\begin{equation}
\begin{aligned}\label{eq:solutionpsi0}
W=&\frac{X^2}{2g^2 (2X^2+1)}\left(1+\frac{2 \alpha  g^2 \lambda _1 \left(8 X^8-26 X^6+41 X^4-12 X^2+1\right)}{X^4 \left(8 X^4-10 X^2+5\right)}\right)\, ,\\
G=&\frac{\left(1-x^2\right) \left(1-4 X^2\right)}{2 X^2+1}\left(1-\frac{32 \alpha  g^2 \lambda _1 \left(2 X^6-3 X^2+1\right)}{X^2 \left(8 X^4-10 X^2+5\right)}\right)\, ,\\
   B=&\left(\frac{X^2}{g(2 X^2+1)}\right)^2\left(1+\frac{2 \alpha  g^2 \lambda _1 \left(X^2-1\right)^2 \left(40 X^4-2 X^2+1\right)}{X^4 \left(8 X^4-10 X^2+5\right)}\right)\, ,\\
   J=&\frac{2 g^3 x}{X^3} \sqrt{1-2 X^2} \left(2 X^2+1\right)^2\left(1-\frac{\alpha  g^2 \lambda _1 x \left(120 X^8-134 X^6+31 X^4+4 X^2+3\right)}{X^4 \left(8 X^4-10 X^2+5\right)}\right)\, ,\\
   \hat{A}_{1}=&0\,  ,\\
\hat{A}_{2}=&-\frac{4 \alpha  g \lambda _1 \sqrt{3-6 X^2}}{X(2 X^2+1)}\, ,
\end{aligned}
\end{equation}
so a non-trivial $\hat{A}_{2}$ is generated. The reason for this is the gravitational Chern-Simons term, which acts as a source in Maxwell equation, as $d\star F\sim \alpha R\wedge R$. Therefore, gravity --- and more precisely, angular momentum --- can generate an electric charge. We obtain the precise value of this charge in the next section.

\subsection{Supersymmetric solutions with $J_{1}\neq J_{2}$  }\label{sec:susysol}
When $J_{1}\neq J_{2}$, the equations of motion are much more involved and we have not been able to find an exact solution. 
However, we can find an approximate solution when the difference between angular momenta is small by performing an expansion in powers of $(J_{1}-J_{2})$ around the solution with equal angular momenta. To simplify things further, we also restrict ourselves to the more holographically interesting supersymmetric case. 

Let us note that, in the two-derivative theory, we identified supersymmetry with the constraint \req{susyXYZ}, which in turn implied that the chemical potentials satisfy the relation \req{eq:susyconstr}. In general, a solution is supersymmetric when it allows for a non-zero Killing spinor. Now, the Killing spinor equations are in general modified by the higher-derivative corrections, and hence one should compute what the effect is on the supersymmetry conditions. Following \cite{Cassani:2022lrk}, we assume that the constraint \req{eq:susyconstr} is not modified by the higher-derivative corrections and we take this to be the defining property of a supersymmetric solution at subleading order in $\alpha$. Since we are working in the grand-canonical ensemble, we make sure that the correction to the two-derivative supersymmetric solution still satisfies \req{eq:susyconstr} and hence is still supersymmetric by setting
\begin{equation}
Z=1-X-Y\, .
\end{equation}
In order to find a solution, our goal is to expand the higher-derivative corrections in \req{funcexpan} in a power series of $(Y-X)$, since this measures the difference between the two angular momenta.

There is another key property of the solution that makes the expansion less burdensome. We observe that, whenever the variable $x$ appears nontrivially\footnote{With this we mean that there is an additional $x$-dependence with respect to the $X=Y$ case.} in the two-derivative solution \req{eq:solution2der}, it is always multiplied by $(Y-X)$. Thus, when we expand this solution in powers of $(Y-X)$, the $n$-th power contains, at most, $n$ powers of $x$. It is very natural to assume that the higher-derivative corrections will have the same structure. Taking this into account, we expand the correction to the functions in \req{funcexpan} as
\begin{equation}\label{funcexpand3}
\begin{aligned}
W^{(1)}(x)=&\sum_{n=0}^{\infty}(Y-X)^n\sum_{k=0}^{n}W_{n,k} x^k\, ,\quad &
G^{(1)}(x)=&(1-x^2)\sum_{n=0}^{\infty}(Y-X)^n\sum_{k=0}^{n}G_{n,k} x^k\, ,\\
B^{(1)}(x)=&\sum_{n=0}^{\infty}(Y-X)^n\sum_{k=0}^{n}B_{n,k} x^k\, ,\quad &
j^{(1)}(x)=&\sum_{n=0}^{\infty}(Y-X)^n\sum_{k=0}^{n}j_{n,k} x^k\, ,\\
\zeta^{(1)}(x)=&\sum_{n=0}^{\infty}(Y-X)^n\sum_{k=0}^{n}\zeta_{n,k} x^k\, ,\quad &
\hat{A}_{2}^{(1)}(x)=&\sum_{n=0}^{\infty}(Y-X)^n\sum_{k=0}^{n}\hat{A}_{2, n,k} x^k\, ,
\end{aligned}
\end{equation}
where $W_{n,k}$, $G_{n,k}$, $B_{n,k}$, $j_{n,k}$, $\zeta_{n,k}$ and $\hat{A}_{2,n,k}$ are constant coefficients. By plugging this ansatz into the equations of motion, expanding again in powers of $(Y-X)$, and collecting the terms with different powers of $x$,\footnote{Observe that we do not perform a series expansion in $x$, we keep all the $x$-dependence at each order of $(Y-X)$. This dependence on $x$ is automatically polynomial. } we obtain a system of algebraic equations for these coefficients. We check that this system indeed solves the equations of motion order by order, hence validating the form of the ansatz \req{funcexpand3}.

The solution in fact contains free coefficients that must be fixed by imposing the conditions \req{bdycond}. To this end, we first obtain the functions $J(x)$ and $\hat{A}_{1}(x)$  by plugging \req{funcexpan} with \req{funcexpand3} into \req{eq:JAprime}, and expanding again in $(Y-X)$. In this case, the expansion of these functions takes the form 
\begin{equation}\label{funcexpandJA}
J^{(1)}(x)=\sum_{n=0}^{\infty}(Y-X)^n\sum_{k=0}^{n}J_{n,k} x^{k+1}\, ,\quad \hat{A}_{1}^{(1)}(x)=\sum_{n=0}^{\infty}(Y-X)^n\sum_{k=0}^{n}\hat{A}_{1, n,k} x^{k+1}\, ,
\end{equation}
and the coefficients $J_{n,k}$, $\hat{A}_{1,n,k}$ are determined by the ones in \req{funcexpand3}.
This allows us to evaluate the correction to the potentials in \req{correctionpotentials}, which also take the form of a series expansion, and hence \req{bdycond} provides three constraints at every order of $(Y-X)$. When these constraints are imposed together with the equations of motion, we find that the solution of the form \req{funcexpand3} is unique.  We have been able to obtain the solution to order $(Y-X)^6$ and we provide the corresponding coefficients in the ancillary Mathematica notebook. For the sake of illustration, we show the solution to first order in $(Y-X)$ in Appendix~\ref{app:solution}. 

We have attempted to find a pattern that would allow us to sum the whole series expansion and obtain the exact solution. In particular, we tried to fit the solution to a rational function of $x$, in analogy with the two-derivative solution \req{eq:solution2der}. However, the expressions are too complicated and we could not find a simple pattern. Nevertheless,  we will see in Section \ref{subsec:resum} that there is a way in which this can be done in the case of the thermodynamic quantities.

With the solution at hand we can compute right away the area of the horizon by applying \req{eq:areagen}. We get
\begin{equation}
\begin{aligned}
\mathcal{A}=&\frac{2\pi ^2 (X+Y) \sqrt{X+Y-X^2-X Y-Y^2}}{g^3 (1-2 X-Y) (1-2 Y-X)}\Big[1+\alpha g^2 \lambda_{1}\delta\mathcal{A}\Big]\, ,
\end{aligned}
\end{equation}
where, to second order in $(Y-X)$, $\delta \mathcal{A}$ reads
\begin{equation}
\begin{aligned}
\delta\mathcal{A}=&-\frac{-180 X^3+177 X^2-32 X+9}{X(9 X^2-12 X+1)}-\frac{9 (Y-X)  \left(63 X^4+24 X^3-50 X^2+24 X-1\right)}{2X^2 \left(9 X^2-12 X+1\right)^2}\\
&+\frac{(Y-X)^2}{12 (X-1)^2 X^3 \left(3 X^2-1\right) \left(9 X^2-12 X+1\right)^3} \Big(229635 X^{11}-946971 X^{10}+1665765 X^9\\
&-1705617 X^8+1153926 X^7-495990 X^6+67410 X^5+76662 X^4-56937 X^3\\
&+14737 X^2-1047 X+27\Big)+\mathcal{O}\left((Y-X)^3\right)\, .
\end{aligned}
\end{equation}
Although this is again terribly complicated to guess a pattern, we will find a way to sum the whole series as we elaborate on in Section \ref{subsec:resum}.

\section{Thermodynamics}
\label{sec:thermo}

\subsection{Electric charge: ambiguity due to the Lorentz-Chern-Simons terms}\label{subsec:Qamb}

The formulas for the electric charge \eqref{eq:Qint}, angular momenta \eqref{eq:Jint} and entropy \eqref{eq:Waldentropy} present gauge ambiguities due to the presence of Chern-Simons terms. However, these ambiguities are fixed by working in the regular gauge that we are utilizing, \textit{i.e.}, imposing regularity at the horizon. However, in the formula for the electric charge there is also a Lorentz-Chern-Simons three-form that introduces frame ambiguities. These are harder to fix, since there is not a priori a canonical choice of frame. Instead, there are infinitely many regular frames, which are related by ``large'' Lorentz transformations. In our case, since the spacetime is a twisted product AdS$_{2}\times \mathbb{S}^{3}$, it is natural to restrict to the rotations of frame generated by the sphere $\mathbb{S}^{3}$, leading to a discrete set of different notions of charge --- one for each homotopy class of  $\mathbb{S}^{3}$. 
Each of these is, in principle, a valid definition of electric charge. However, only one particular charge (modulo an additive constant independent of the parameters of the solution) enters in the first law of thermodynamics. Then, the question is whether we can find a frame in which the electric charge is the thermodynamically relevant charge. 

With this goal, we consider a family of frames $\tilde{e}^{a}$ related to \req{framedef} according to
\begin{equation}\label{frame2}
\begin{aligned}
\tilde{e}^{0}&=e^{0}\, ,\\
\tilde{e}^{1}&=e^{1}\, ,\\
\tilde{e}^{2}&=\cos(n\phi_{1}+m\phi_{2})e^{2}+\sin(n\phi_{1}+m\phi_{2})e^{3}\, ,\\
\tilde{e}^{3}&=-\sin(n\phi_{1}+m\phi_{2})e^{2}+\cos(n\phi_{1}+m\phi_{2})e^{3}\, ,\\
\tilde{e}^{4}&=e^{4}\, ,
\end{aligned}
\end{equation}
where $\phi_{1}$ and $\phi_{2}$ are the angular coordinates introduced in \req{eq:angles} and $n,m$ must be integers in order to ensure the regularity of the transformation. Although this is clearly not the most general choice of frame, we show below that this is indeed enough for our purposes. 
Due to the frame transformation, the Lorentz-Chern-Simons three-form introduces a non-trivial contribution to the charge, which reads
\begin{equation}
\tilde{\Omega}_{\rm LCS}=\Omega_{\rm LCS}+\mathcal{W}\,,
\end{equation}
where 
\begin{equation}
\mathcal{W}=dy_{1}\wedge dy_{2}\wedge d\left[-\sigma_{1} B^{3/2} J'+\frac{\sigma_{2}}{B^{3/2} W^2}  \left(B W \left(G'+B^2 J W J'\right)-G
   \left(W B'+B W'\right)\right)\right]\, ,
\end{equation}
and 
\begin{equation}
\begin{aligned}
\sigma_{1}&=\frac{1}{2 (J(-1)-J(1))}\left[\frac{m J(-1) G'(1)}{\sqrt{B(1)} W(1)}-\frac{n J(1)G'(-1)}{\sqrt{B(-1)} W(-1)}\right]\, ,\\
\sigma_{2}&=\frac{1}{2 (J(-1)-J(1))}\left[\frac{m G'(1)}{\sqrt{B(1)} W(1)}-\frac{n G'(-1)}{\sqrt{B(-1)} W(-1)}\right]\, .
   \end{aligned}
\end{equation}
These coefficients come  from the relation \req{eq:angles} between the angles $\phi_{i}$ and the $y_{i}$ coordinates. When we integrate the charge, we get the following difference due to the frame change
\begin{equation}
\Delta Q=\frac{3\alpha \lambda_{1}}{16\sqrt{3}\pi G_{5}}\int \mathcal{W}\, .
\end{equation}
At first order in $\alpha$, we can just evaluate this term at the two-derivative solution, and we get the following explicit result for the ambiguous contribution to the charge as a function of $n$ and $m$, 

\begin{equation}\label{deltaQnm}
\Delta Q_{n,m}=\frac{\alpha \lambda_{1} \pi}{2\sqrt{3}\pi G_{5}}\left[n f(X,Y,Z)+m f(Y,X,Z)\right]\, ,
\end{equation}
where 

\begin{equation}\label{fcharge}
\begin{aligned}
	f(X,Y,Z)=&\frac{1}{\left(Y^2-Z^2\right)
   \left(-1-X^2+Y^2+Z^2\right)^3}\bigg[X^7 Y+X^5 Y \left(5-3 Y^2+Z^2\right)\\
   &-X Y
   \left(-1+Y^2+Z^2\right)^2 \left(1+Y^2+Z^2\right)+X^6 \left(-Y^2+Z
   (2+Z)\right)\\
   &+\left(-1+Y^2+Z^2\right)^2 \left(Y^4+Z^2-Z^4-Y^2 (1+2
   Z)\right)+X^4 \left(3 Y^4+Y^2 (-3+10 Z)\right.\\
   &\left.-3 Z^2
   \left(-1+Z^2\right)\right)+X^3 Y \left(-5+3 Y^4+6 Z^2-Z^4+2 Y^2
   \left(1+5 Z^2\right)\right)\\
   &+X^2 \left(-3 Y^6+Y^4 \left(6+6 Z-3
   Z^2\right)+Z (-2+3 Z) \left(-1+Z^2\right)^2\right.\\
   &\left.+Y^2 \left(-3-4 Z+4 Z^3+3
   Z^4\right)\right)\bigg]\, .
\end{aligned}
\end{equation}
Since this is not a constant value, it affects the variations of the charge $dQ_{n,m}$ and therefore it affects the first law. Our hope is that there is a choice of $n$ and $m$ for which $Q_{n,m}$ is the thermodynamic charge. We study this in careful detail below.

\subsection{Equal angular momenta}
By direct evaluation of the integrals \eqref{eq:Qint}, \eqref{eq:Jint} and \eqref{eq:Waldentropy} in the solution \req{eq:solution2derequalJ}, we obtain the electric charge, angular momenta and the entropy of the black holes. The result takes the form
\begin{align}
S&=\frac{\pi ^2 (Z+1) \left(2 X^2-Z^2+Z\right) \sqrt{-2 X^2-Z^2+1}}{4 \sqrt{2} G_{5} g^3 \left(X^2-Z^2\right)^2}\left[1+\alpha g^2(4\lambda_{2}+\lambda_{1}\Delta S)\right]\, ,\\
		J_{i}&=\frac{\pi  X (Z+1) \left(X^2 (6 Z-2)+4 X^4+Z^4+Z^3+Z^2-3Z\right)}{16 G_{5} g^3 \left(X^2-Z^2\right)^3}\left[1+\alpha g^2(4\lambda_{2}+\lambda_{1}\Delta J)\right]\, ,\\
		Q&=\frac{\sqrt{3} \pi  Z (Z+1) \left(1-Z-2 X^2\right)}{8 G_{5} g^2 \left(X^2-Z^2\right)^2}\left[1+\alpha g^2(4\lambda_{2}+\lambda_{1}\Delta Q)\right]\\
	&+\frac{\pi(m+n) \alpha \lambda_{1} \left[(Z-1)^2 \left(X^2 (4 Z+2)+4 X^4+Z^2 \left(Z^2-1\right)\right)-8 X^6\right]}{2 \sqrt{3} G_{5} (Z-1)^3 (Z+1) (Z^2-X^2)}\, , \nonumber
\end{align}
where the expressions of $\Delta S$, $\Delta J$, $\Delta Q$ are somewhat long and hence we provide them in the ancillary Mathematica notebook. Observe that the correction due to $\lambda_{2}$ is universal and it is equivalent to replacing $G_5$ by the effective Newton's constant $G_{\rm eff}$ in \req{Geff}. 

In the case of $Q$, we have made explicit the ambiguous Chern-Simons contribution due to the frame choice, that in this case only depends on the sum of the two integers $n+m$. In order to fix this ambiguity, we evaluate the \textit{near-horizon version} of the first law of black hole mechanics \req{1stlawhorizon}, which should hold as well in the presence of higher-derivative corrections. We get

\begin{equation}
\frac{1}{2\pi}dS-\omega_{1}\, dJ_{1}-\omega_{2}\, dJ_{2}-\psi\, dQ=-(m+n+4) \frac{\pi  \alpha \lambda_{1}}{2 \sqrt{3} G}\psi d[f(X,X,Z)]\, ,
\end{equation}
where $\omega_{i}$ and $\psi$ are the potentials given by \req{potentials2d} evaluated at $Y=X$ --- these receive no corrections since we are working in the grand-canonical ensemble --- and $f$ is the function introduced in \req{fcharge}. Thus, for
\begin{equation}
n+m=-4,
\end{equation}
the first law is satisfied and hence $Q$ becomes the thermodynamic electric charge. 

\subsubsection*{Supersymmetric limit}
In the supersymmetric case, \eqref{susyXYZ2} with $a=b$, \textit{i.e.}, $X=ag/(1+2ag)$, $Z=1/(1+2a g)$, the expressions above simplify dramatically and we get
\begin{equation}
\begin{aligned}
S=&\frac{\pi ^2 a \sqrt{a g (a g+2)}}{g^2 G_{\rm eff} (1-a g)^2}\left[1+\alpha g^2\lambda_{1}\frac{48 (a g+2) (2 a g+1)}{11 a^2 g^2+8 a g-1}\right]\, ,\\
J_{i}=&\frac{\pi  a^2 (a g+3)}{2 g G_{\rm eff} (1-a g)^3}\left[1+\alpha g^2\lambda_{1}\frac{24 \left(8 a^4 g^4+25 a^3 g^3+29 a^2 g^2+9 a g+1\right)}{ag (a g+3) \left(11 a^2 g^2+8 a g-1\right)}\right]\, ,\\ 
Q=&\frac{\sqrt{3} \pi  a}{2 g G_{\rm eff} (1-a g)^2}\left[1+\alpha g^2\lambda_{1}\frac{8 \left(23 a^4 g^4+73 a^3 g^3+57 a^2 g^2+7 a g+2\right)}{3 ag \left(11 a^2 g^2+8 a g-1\right)}\right]\, .
\end{aligned}
\end{equation}
These results match exactly those of \cite{Cassani:2023vsa}, also obtained from the near-horizon geometry. However, as also noticed by \cite{Cassani:2023vsa}, the charge obtained from this procedure differs by a pure constant --- independent of the parameters of the black hole --- from the charge obtained from the on-shell action \cite{Cassani:2022lrk} (the rest of the results are identical). In the next subsection we analyze this discordance in the case of different angular momenta.

\subsubsection*{Zero chemical potential} 
As we noted in Section~\ref{sec:equalJsol}, the solutions with $\psi=0$ actually acquire a nontrivial gauge field on account of the electro-gravitational Chern-Simons term. These solutions even have a nonzero charge, and in fact we get 
\begin{equation}\label{Qzeromu}
Q=-\frac{4 \pi  \alpha \lambda_{1} \left(3-4 X^2-8 X^4\right)}{\sqrt{3} G_{5} \left(1-4 X^2\right)^2}\, ,
\end{equation}
where we have set $n+m=-4$, as before. Observe that, since $Q$ depends on $X$, it cannot be taken to zero by adding a pure constant to it. In fact, it is not zero in any of the frames we have analyzed, so the generation of a non-zero charge is a genuine effect of the higher-derivative corrections. Interestingly, $Q$ is a function of the angular momentum through its dependence on $X$, so it is not a free parameter of the solution. On the other hand, one may also consider solutions with zero charge, but these will necessarily have a nonzero potential. 

\subsection{Supersymmetric solutions with $J_{1}\neq J_{2}$}
Let us now consider the supersymmetric solutions that we have found in Section~\ref{sec:susysol} as an expansion in $(Y-X)$. We can again evaluate the charges and the entropy through the integrals \eqref{eq:Qint}, \eqref{eq:Jint} and \eqref{eq:Waldentropy}. In the case of the electric charge we also add the frame-dependent term  \req{deltaQnm}. In order to make contact with previous literature, we express the result in terms of the parameters $a$ and $b$ by using the relationships \req{susyXYZ2}. Then, we also translate the series expansion in $(Y-X)$ into a series expansion in $(b-a)$. The result, to order $(b-a)^2$ and first order in $\alpha$, reads
\begin{equation}\label{thermocorrect}
\begin{aligned} 
	 S&=\frac{\pi ^2 (a+b) \sqrt{a b +a+b}}{2 G_{\rm eff}  (1-a ) (1- b)}\bigg[1+\alpha \lambda_{1}\delta S\bigg]\, ,  \\
	J_{1}&=\frac{\pi  (a+b) (a b+2a+b)}{4 G_{\rm eff} (1-a)^2 (1-b)}\bigg[1+\alpha \lambda_{1}\delta J_{1}\bigg]\, , \\
	J_{2}&=\frac{\pi  (a+b) (a b+2b+a)}{4 G_{\rm eff}  (1-b)^2 (1-a)}\bigg[1+\alpha \lambda_{1}\delta J_{2}\bigg]\,,\\
	Q&=\frac{\sqrt{3} \pi  (a+b)}{4 G_{\rm eff}  (1-a) (1-b)}\bigg[1+\alpha \lambda_{1}\delta Q_{n,m}\bigg]\, , 
\end{aligned}
\end{equation}
where we have set $g=1$ and

\begin{align}\notag
\delta S&=\frac{48  (a +2) (2 a +1)}{11 a^2+8 a-1}-\frac{72 (b-a) \left(13 a^2 +16 a +7\right)}{\left(11 a^2 +8 a -1\right)^2}\\\notag
&-\frac{2  (b-a)^2}{a^2 (a +1)^2 \left(a^2 +4 a +1\right) \left(11 a^2 +8 a -1\right)^3} \big(33 a^{10}-5520 a^9-33975 a^8 -81460 a^7 \\
&-108482 a^6 -88572 a^5 -43066 a^4 -11036 a^3 -1167 a^2 -4 a +1\big)+\mathcal{O}\left((b-a)^3\right)\, ,
\label{deltaS1}
	\\[10pt] \notag
\delta J_{1}&=\frac{24 \left(8 a^4+25 a^3+29 a^2+9 a+1\right)}{a (a+3) \left(11 a^2+8
   a-1\right)}+\frac{8(b-a)}{a^2 (a+3)^2 \left(a^2+4
   a+1\right) \left(11 a^2+8 a-1\right)^2} \big(88 a^9\\\notag
   &+1189 a^8+5762 a^7+13249 a^6+14422 a^5+7889
   a^4+3094 a^3+895 a^2+74 a-6\big) \\\notag
   &+\frac{2(b-a)^2}{a^3 (a+1)^2 (a+3)^3
   \left(a^2+4 a+1\right)^2 \left(11 a^2+8 a-1\right)^3} \big(2475 a^{16}+73157 a^{15}\\\notag
   &+778745 a^{14}+4365729
   a^{13}+14805057 a^{12}+32337717 a^{11}+47141825 a^{10}\\\notag
   &+47441601
   a^9+34726215 a^8+19894695 a^7+9289651 a^6+3306235 a^5+754083
   a^4\\
   &+76383 a^3-2157 a^2-605 a+42\big) +\mathcal{O}\left((b-a)^3\right)\, ,\\[10pt]
   \label{deltaJ1}
\notag
\delta Q_{n,m}&=\frac{2 \left(2+17 a+63 a^2+85 a^3+49 a^4\right)}{a \left(-1+8 a+11
   a^2\right)}+\frac{\left(2-7 a+5 a^2\right) (m+n)}{6 a}\\\notag
   &+(b-a)\Bigg[\frac{2-32 a-265 a^2-544 a^3-160 a^4+784 a^5+539 a^6}{a^2 \left(-1+8
   a+11 a^2\right)^2}\\\notag
   &-\frac{\left(1+a-5 a^2-3 a^3\right) m+\left(1+a-2
   a^3\right) n}{6 a^2 (1+a)}\Bigg]\\\notag
   &+(b-a)^2\Bigg[\frac{1}{6
   a^3 (1+a)^2 \left(1+4 a+a^2\right) \left(-1+8 a+11
   a^2\right)^3}\big( 6-111 a+93 a^2+14383 a^3\\\notag
   &+95277 a^4+282930 a^5+497994 a^6+614982
   a^7+522456 a^8+208221 a^9-50727 a^{10}\\\notag
   &-70485 a^{11}-15499 a^{12}\big)+\frac{\left(2+3 a+3 a^2-5 a^4\right)
   (n+m)-a^4(11m+3n)}{24 a^3 (1+a)^2}\Bigg]\\
   &+\mathcal{O}\left((b-a)^3\right)\, .
\end{align}
and, as usual, $\delta J_{2}=\delta J_{1}(a\leftrightarrow b)$. We observe that the results for $S$ and $J_{i}$ coincide with those of \cite{Cassani:2022lrk} when expanded in $(b-a)$, which represents a highly nontrivial check of our computations. On the other hand, we have to determine for which value of $(n,m)$ the electric charge is the thermodynamic one. For this charge, the near-horizon version of the first law of thermodynamics should hold. By plugging \req{thermocorrect} and \req{potentials2d} in the supersymmetric case into \req{1stlawhorizon}, we obtain\footnote{Naturally, we obtain the result as a series expansion in $(b-a)$, but we verify that it corresponds to the expansion of this expression.}

\begin{equation}
\frac{1}{2\pi}dS-\omega_{1}\, dJ_{1}-\omega_{2}\, dJ_{2}-\psi\, dQ=-\frac{\pi  \alpha \lambda_{1}}{2 \sqrt{3} G_{5}}\psi\, d\left[(n+2)f^{*}(a,b)+(m+2)f^{*}(b,a)\right]\, ,
\end{equation}
where
\begin{equation}
f^{*}(a,b)\equiv f(X^{*},Y^{*},Z^{*})=\frac{3 a^2 b+2 a b^2+a b+2 b^2-a-b}{(a+1) (b-1) (a+b)}\, .
\end{equation}
Therefore, the first law is satisfied for
\begin{equation}
n=m=-2\, ,
\end{equation}
and thus the frame \req{frame2} is fully fixed by this criterion. 
For these values of $n$ and $m$, the charge can be rewritten as 
\begin{equation} \label{eq:Qwithcorrections}
Q=\frac{\sqrt{3} \pi  (a+b)}{4 G_5 (1-a) (1-b)}\bigg[1+4\alpha \lambda_{2}+\alpha \lambda_{1}\delta Q\bigg]-\frac{2\pi \alpha \lambda_{1}}{\sqrt{3}G_5}\, ,
\end{equation}
with 
\begin{equation}
\begin{aligned}
\delta Q&=\frac{4 \left(19 a^4+44 a^3+36 a^2+8 a+1\right)}{a \left(11 a^2+8 a-1\right)}\\
&+\frac{(b-a)\left(418 a^6+608 a^5-202 a^4-528 a^3-266 a^2-32 a+2\right)}{a^2 \left(11 a^2+8 a-1\right)^2}\\
&+\frac{(b-a)^2}{3 a^3 (a+1)^2 \left(a^2+4 a+1\right) \left(11 a^2+8 a-1\right)^3}\big(-5753 a^{12}-20904 a^{11}+7026 a^{10}\\
&+129304 a^9+260413 a^8+301360 a^7+250612 a^6+143104 a^5+47449 a^4+7128 a^3\\
&+58 a^2-56 a+3\big) +\mathcal{O}\left((b-a)^3\right)\, .
\label{deltaQ1}
\end{aligned}
\end{equation}
In this case, the electric charge coincides with the one in \cite{Cassani:2022lrk} except for the last term in \eqref{eq:Qwithcorrections}. Indeed, we have
\begin{equation}\label{chargediff}
Q-Q_{ \text{\cite{Cassani:2022lrk}}}=-\frac{2\pi \alpha \lambda_{1}}{\sqrt{3}G_5}\, .
\end{equation}
This constant shift obviously does not affect the first law and it coincides with the shift obtained in \cite{Cassani:2023vsa} from the near-horizon geometry of the solution with $a=b$. It would be interesting to look for a different frame --- outside of the family of frames \req{frame2} ---  in which one can further remove this constant. However, given that this constant is a pure number --- does not depend on the solution --- this mismatch is not too worrisome. 

\subsection{Exact expressions}\label{subsec:resum}
With the equations \req{deltaS1}, \req{deltaJ1}, \req{deltaQ1} at hand it is quite challenging to study the relation between the entropy and the charges, due the complexity of these expressions. However, most of this complexity is due to our requirement of working in the grand-canonical ensemble. Instead, we can do a relabeling of the $a$ and $b$ parameters as 
\begin{equation}
(a,b)\rightarrow (a+\alpha \lambda_{1}\delta a, b+\alpha \lambda_{1} \delta b)\, ,
\end{equation}
which has the effect of changing the corrections $\delta S$, $\delta J_{i}$ and $\delta Q$ in \req{thermocorrect} and \req{deltaQ1}. We can use this freedom to cancel for instance the $\lambda_{1}$ corrections to the expressions of the angular momenta $J_{i}$. 
Thus, in this ``fixed angular momenta ensemble'' our results read
\begin{equation}
\begin{aligned}
S=&\frac{\pi ^2 (a+b) \sqrt{a b +a+b}}{2 G_{\rm eff}  (1-a ) (1- b)}\bigg[1+\alpha \lambda_{1}\delta \tilde S\bigg]\, ,\\
J_{1}=&\frac{\pi  (a+b) (a b+2a+b)}{4 G_{\rm eff} (1-a)^2 (1-b)}\, ,\\
J_{2}=&\frac{\pi  (a+b) (a b+2b+a)}{4 G_{\rm eff} (1-b)^2 (1-a)}\, ,\\
Q=&\frac{\sqrt{3} \pi  (a+b)}{4 G_{\rm eff} (1-a) (1-b)}\bigg[1+\alpha \lambda_{1}\delta \tilde Q\bigg]\, ,
\end{aligned}
\end{equation}
where $\delta \tilde S$ and $\delta \tilde Q$ are now given by

\begin{equation}
\begin{aligned}
\delta \tilde S&=\frac{12 (a+1)}{a (a+2)}-\frac{6 \left(a^2+2 a+2\right)(b-a)}{a^2 (a+2)^2}+\frac{\left(a^5+9 a^4+21 a^3+26 a^2+24 a+12\right)(b-a)^2}{a^3 (a+1) (a+2)^3}\\
&-\frac{\left(2 a^7+20 a^6+62 a^5+91 a^4+88 a^3+88 a^2+72 a+24\right) (b-a)^3}{2 a^4 (a+1)^2 (a+2)^4} +\mathcal{O}\left((b-a)^4\right)\, ,\\
\delta \tilde Q&=\frac{4 \left(2 a^2+5 a-1\right)}{3 a}+\frac{2 \left(2 a^2+1\right) (b-a)}{3 a^2}
-\frac{\left(3 a^3+5 a^2+2 a+1\right) (b-a)^2}{3 a^3 (a+1)}\\
&+\frac{\left(3 a^4+10 a^3+9 a^2+4 a+1\right) (b-a)^3}{6 a^4 (a+1)^2}+\mathcal{O}\left((b-a)^4\right)\, .
\end{aligned}
\end{equation}
These expressions are now simple enough that we can guess the exact result. We find that the simplest rational functions of $a$ and $b$ --- with the lowest order numerator and denominator --- leading to the expansions above are
\begin{equation}\label{eq:deltaSQ}
\begin{aligned}
\delta \tilde S=&\frac{4 \left[a b \left((a-b)^2+5(a+b)+18\right)+a^3+b^3+3 (a^2+b^2)+6(a+b)\right]}{(a+b) (a+b+2) (a b+a+b)}\, ,\\
\delta \tilde Q=&\frac{8 \left[a b \left(a^2+6 a b+b^2+13(a+b)+10\right)+a^3+b^3+3 (a^2+b^2)-2(a+b)\right]}{3 (a+b)^2 (a+b+2)}\, .
\end{aligned}
\end{equation}
We then check that these expressions also capture higher-order terms in the $(b-a)$ expansion --- we checked up to order $(b-a)^6$ --- and therefore this result must be exact.  With these expressions we can now study the dependence of the entropy on the charges as well as the non-linear constraint satisfied by these. One can directly check that the following expression 
\begin{equation}
S=\pi\sqrt{4Q^2-\frac{\pi}{G_{\rm eff}}(J_{1}+J_{2})+\frac{8\pi\alpha\lambda_{1}}{G_{\rm eff}}\left[\frac{2Q}{\sqrt{3}}+\frac{\pi(J_{1}-J_{2})^2}{16G_{\rm eff}Q^2/3-\pi(J_{1}+J_{2})}\right]}\,,
\end{equation}
holds up to $\mathcal{O}(\alpha^2)$ terms. This coincides with the result of \cite{Bobev:2022bjm, Cassani:2022lrk} except for the linear term in the charge, which comes precisely from the difference in the definition of the charges \req{chargediff}. 
Now, in order to derive the non-linear constraint between the charges, the easiest way consists in evaluating the zeroth-order constraint \req{eq:constr0} to obtain how much it fails to be satisfied. Since in this parametrization only the electric charge is modified, we get
\begin{equation}\label{constrcorr}
\left[2\sqrt{3}Q+\frac{\pi}{2G_{\rm eff}}\right]\left[4Q^2-\frac{\pi}{G_{\rm eff}}(J_1+J_2)\right]-\left(\frac{2Q}{\sqrt{3}}\right)^3-\frac{2\pi}{G_{\rm eff}}J_1J_2=\frac{\alpha\lambda_{1}}{G_{\rm eff}^2}\Sigma+\mathcal{O}(\alpha^2)\, , 
\end{equation}
with
\begin{equation}
\Sigma=\frac{2 Q \delta\tilde Q}{3G_{\rm eff}} \left(32 \sqrt{3} G_{\rm eff} Q^2-3 \pi  \sqrt{3} (J_1+J_2)+6 \pi  Q\right)\, .
\end{equation}
Thus, in order to obtain an explicit constraint we only need to express $ \delta\tilde Q$ as function of the charges. For this it suffices to use the two-derivative relations between $(a,b)$ and $(J_{i}, Q)$, since the right-hand side of \req{constrcorr} is already proportional to $\alpha$. 
After some massaging, we find that $\Sigma$ can be written as
\begin{equation}
\begin{aligned}
\Sigma=&48 \pi G_{\rm eff} J_{1}J_{2}+4\pi^2  \left(J_1+J_2\right)-\frac{8 \pi  Q}{\sqrt{3}} \left(\pi -6 G_{\rm eff} \left(J_1+J_2\right)\right)-\frac{80}{3} \pi  G_{\rm eff} Q^2\\
&+\frac{4 \pi ^2 \left(J_1-J_2\right){}^2 \left(16 \sqrt{3} G_{\rm eff} Q+3 \pi \right)}{3 \pi  \left(J_1+J_2\right)-16 G_{\rm eff} Q^2}\, .
\end{aligned}
\end{equation}

Finally, we can rewrite the entropy and the constraint in terms of the central charges of the dual CFT \req{acdef}. We have
\begin{equation}
S=\pi\sqrt{4Q^2-8\mathfrak{a}(J_{1}+J_{2})-8(\mathfrak{a}-\mathfrak{c})\left[\frac{2Q}{\sqrt{3}}+\frac{\mathfrak{a}(J_{1}-J_{2})^2}{2Q^2/3-\mathfrak{a}(J_{1}+J_{2})}\right]}\, ,
\end{equation}
which again matches the results of \cite{Bobev:2022bjm, Cassani:2022lrk} --- and hence the field theory prediction --- once we take into account \req{chargediff}.
On the other hand, the non-linear constraint can be expressed as 
\begin{equation}\label{constrcorr3}
\begin{aligned}
\left[\sqrt{3} Q+2 \mathfrak{a}\right] \left[Q^2-2 \mathfrak{c} \left(J_1+J_2\right)\right]=&\frac{Q^3}{3 \sqrt{3}}+2 (3 \mathfrak{c}-2 \mathfrak{a}) J_1 J_2+\frac{2}{3}(\mathfrak{a}-\mathfrak{c})\Bigg[4 \sqrt{3} \mathfrak{a} Q+5Q^2\\
&+\frac{6 \mathfrak{a} \left(3 \mathfrak{a}+2 \sqrt{3} Q\right)
   \left(J_1-J_2\right){}^2}{2 Q^2-3 \mathfrak{a} \left(J_1+J_2\right)}\Bigg]\, .
\end{aligned}
\end{equation}
This again coincides with the results of \cite{Cassani:2022lrk} taking into account \req{chargediff} once more. 
\section{Conclusions  }
\label{sec:conclusions}
We have studied extremal black holes in AdS$_5$ supergravity with four-derivative corrections by analyzing their near-horizon geometries. We introduced a novel parametrization of these solutions that allowed us to write them explicitly and to solve the corrected Maxwell and Einstein equations. Then, we were able to compute the charges of the black holes as integrals over the horizon, thus allowing us to study the different thermodynamic relations even for different angular momenta.  We observed that the entropy, charges and near-horizon potentials satisfy a near-horizon version of the first-law of thermodynamics \req{1stlawhorizon} which holds as well in the presence of corrections --- see \cite{Hajian:2013lna} for a general analysis of the laws of near-horizon black hole mechanics. This allowed us to identify the near-horizon potentials as fugacities: the derivatives of the actual potentials with respect to the temperature at $T=0$ --- see \req{eq:potdef2}. 

Our results for the thermodynamic quantities in the supersymmetric case match those of \cite{Bobev:2022bjm,Cassani:2022lrk} obtained from evaluation of the on-shell action following the Reall-Santos method \cite{Reall:2019sah}, which does not require solving any equations of motion. This agreement serves as a nontrivial check of both approaches, but nevertheless our computation reveals a number of nuances. 

First, the need to solve the equations of motion makes the computation considerably involved in the case of $J_{1}\neq J_{2}$ and in fact we have not been able to obtain an exact solution. Rather, we had to restrict ourselves to a solution in the form of a series expansion in $(J_{1}-J_{2})$. It would be interesting to obtain a closed form solution but our results indicate that this would be rather complicated as we were not able to find a simple pattern in the series. On the other hand, we did manage to sum up the series expansion for the thermodynamic quantities by working in a fixed angular momentum ensemble --- see \req{eq:deltaSQ} --- which makes the expressions much simpler.

Second, the computation of the electric charge at the near-horizon region is ambiguous due to the presence of the Lorentz-Chern-Simons three-form in the charge integral, whose contribution depends on the frame. We have been able go around this issue by looking for a frame in which the electric charge satisfies the first law of black hole mechanics.  We stress that in general, and unlike what is stated in \cite{Cassani:2023vsa}, a change of frame modifies the charge by a term that depends on the parameters of the solution --- see \req{fcharge} --- and not just by a pure constant.\footnote{Note that the near-horizon geometry is not a direct product AdS$_{2}\times \mathbb{S}^3$ but a fibered product, so one cannot directly apply arguments of 3d gravity to analyze the ambiguities in the Lorentz-Chern-Simons three-form \cite{Witten:2007kt}.} Thus, only in one frame (or restricted family of frames) the charge defined in this way satisfies the first law. 
However, our analysis was not exhaustive and it leaves open the question of adding a pure constant --- independent of the parameters of the solution --- to the charge. In fact, our charge disagrees with the one of \cite{Bobev:2022bjm,Cassani:2022lrk} by a pure constant. The same constant was found in \cite{Cassani:2023vsa} for near-horizon geometries with equal angular momenta in the frame given by the left-invariant Maurer-Cartan forms of $\mathrm{SU}(2)$, and here we showed that it persists unchanged for $J_{1}\neq J_{2}$. 
It would be interesting to characterize all the possible frame transformations and their effect on the charge and to check if one can remove this constant shift in another frame.  More importantly, we would like to understand what makes special the frame \req{frame2} with $n=m=-2$, in which the electric charge is the correct one for thermodynamics. In any case, this shows the difficulty in finding a first-principles interpretation of the charge derived from the on shell action in \cite{Cassani:2022lrk,Bobev:2022bjm}.

Despite these struggles, the analysis of near-horizon geometries does have an advantage over the on-shell action computation: it allows for a direct generalization to subleading higher-derivative corrections. Indeed, the methods and techniques presented here can be extended quite straightforwardly to compute even higher-order corrections. The only real complication would be the increasing length of the equations, but there would not be new obstructions. On the other hand, in order to compute second-order corrections from the on-shell action, one has to obtain the first-order corrections of the full solution \cite{Ma:2023qqj} --- not only the near-horizon geometry. In view of our results here, obtaining the full solution would be very challenging, hence making this approach probably less accessible than the near-horizon analysis.   

On a different note, our analysis was not restricted to supersymmetric solutions, since the space of extremal solutions is larger, although the expressions get substantially longer, even with equal angular momenta.  A particularly interesting phenomenon we observed is that, due to the electro-gravitational Chern-Simons term, gravity sources electromagnetism. In particular, angular momentum sources electric charge.  For instance, the extremal Myers-Perry-AdS solution, which is a member of the family of solutions we are analyzing, necessarily acquires a nontrivial gauge field and it gets a non-zero electric charge given by \req{Qzeromu} (in the case of equal angular momenta). In fact, one cannot have rotating solutions with both zero charge and zero electrostatic potential. It would be interesting to understand the meaning of this effect for the dual field theory, which  must certainly be related to the  mixed $\langle T T \mathcal{R}\rangle$ anomaly \cite{Hanaki:2006pj}. 

There are several interesting avenues that are natural to pursue regarding this work. Our method of studying the near-horizon extremal geometry can be extended to study six-derivative corrections. For that, one must first derive the form of the Lagrangian, which involves starting from the off-shell action formalism, integrating out all auxiliary fields and implementing field redefinitions to find a hopefully relatively compact six-derivative action, see for example \cite{Zucker:1999ej, deWit:2006gn,Hanaki:2006pj,Cremonini:2008tw,Bergshoeff:2011xn,Coomans:2012cf,Ozkan:2013nwa,Baggio:2014hua, Bobev:2021qxx, Cassani:2022lrk,Liu:2022sew}. The difficulty of the analysis of near-horizon geometries in that theory would arise in the length of the computations but it should be in fact feasible. It would then be interesting to match the results with the subleading corrections in the field theory dual.

On the other hand, we can also consider additional matter fields. To the best of our knowledge, higher derivative corrections have not been studied in the context of five-dimensional gauged supergravity with several vector multiplets. This is a rather challenging endeavor as the solutions now involve the metric, gauge potentials and scalars. For example, this would correspond to corrections of the supersymmetric Kunduri-Lucietti-Reall black hole \cite{Kunduri:2006ek}.

Finally, our novel $(X,Y,Z)$ parametrization of extremal AdS$_5$ black holes seems to be a very natural form of expressing these solutions, and it has an interesting geometric interpretation as depicted in Figure~\ref{fig:spaceofsolutions}. This parametrization leads to a compact and explicit way of writing the solution and to a very appealing form of the supersymmetric condition \req{susyXYZ}. 
It would be interesting to see if this kind of parametrization can be generalized to other extremal AdS black holes \cite{Chong:2004na, Chow:2008ip, Wu:2011gq, Bobev:2023bxl}. There are numerous approaches in literature that require the study of extremal solutions and finding such a parametrization for diverse dimensions would be quite advantageous.

\section*{Acknowledgments}
We thank Nikolay Bobev, Davide Cassani, Alejandro Ruip\'erez, Enrico Turetta and  Annelien Vekemans for useful discussions. The work of PAC received the support of a fellowship from “la Caixa” Foundation (ID 100010434) with code LCF/BQ/PI23/11970032. MD is supported in part by the Odysseus grant (G0F9516N Odysseus) as well as from the Postdoctoral Fellows of the Research Foundation - Flanders grant (1235324N).
\appendix

\section{Higher-derivative corrections  }\label{app:solution}
In this appendix, we present the explicit expressions for the higher derivative corrections found in the limit where the angular momenta are equal with no supersymmetry imposed in Section~\ref{appsec:equalJ} and in the supersymmetric case with unequal angular momenta in Section~\ref{appsec:unequalJ}. We note that the results are written in terms of the $(X,Y,Z)$ parametrization.
\subsection{Equal $J$ solution} \label{appsec:equalJ}
We write the various corrections for the $J_1=J_2$ solution. We find that $\delta J = \delta \hat{A}_1$ while the remaining corrections are
\begin{align}
\delta W&=\frac{\lambda_{1}}{\mathcal{D}}\Big[57344 X^{18}-2048 X^{16} \left(92-79 Z+5 Z^2\right)-512 X^{14}
   \left(-507+878 Z-514 Z^2\dvv
   +22 Z^3+65 Z^4\right)+2 (-1+Z)^7 Z^5
   \left(26+39 Z+Z^2-18 Z^3-4 Z^4+3 Z^5+Z^6\right)\dv
   -64 X^{12}
   \left(2997-7610 Z+9159 Z^2-4436 Z^3-1717 Z^4+1550 Z^5+57
   Z^6\right)\dv
   -32 X^{10} (-1+Z)^2 \left(-2423+3006 Z-7319 Z^2-1012
   Z^3+3803 Z^4+502 Z^5+115 Z^6\right)\dv
   -32 X^8 (-1+Z)^3 \left(-461+329
   Z-3408 Z^2-596 Z^3+1995 Z^4+45 Z^5+58 Z^6+70 Z^7\right)\dv
   +16 X^6 (-1+Z)^4 \left(21+122 Z+1338 Z^2-375
   Z^3-1157 Z^4+656 Z^5+268 Z^6-11 Z^7\dvv
   +50 Z^8\right)
   +4 X^4 (-1+Z)^5
   \left(-60+118 Z-200 Z^2-1488 Z^3-569 Z^4+1225 Z^5+94 Z^6\dvv-190 Z^7+207
   Z^8+63 Z^9\right)
   -2 X^2
   (-1+Z)^6 Z^2 \left(360+518 Z-179 Z^2-490 Z^3+199 Z^4\dvv+322 Z^5
   +27Z^6+2 Z^7+9 Z^8\right)\Big]\, ,
\end{align}

\begin{align}
\delta G&=\frac{2 \lambda_{1}(1-Z^2)}{\mathcal{D}} \Big[39936 X^{16}+3072 X^{14} \left(-43+34 Z+9
   Z^2\right)+256 X^{12} \left(721-1144 Z\dvv
   +156 Z^2+284 Z^3+7
   Z^4\right)+128 X^{10} \left(-1087+2545 Z-1441 Z^2-574 Z^3+587 Z^4\dvv
   -35Z^5 +5 Z^6\right)-16 X^8 (-1+Z)^2 \left(-3655+3845 Z-393 Z^2-2606
   Z^3+1007 Z^4\dvv
   +385 Z^5+25 Z^6\right)+(-1+Z)^6 Z^3 \left(54+27 Z-52
   Z^2-26 Z^3+22 Z^4+11 Z^5+8 Z^6\dvv
   +4 Z^7\right)-8 X^6 (-1+Z)^3
   \left(-1578+1219 Z-1531 Z^2-2356 Z^3+1320 Z^4+803 Z^5\dvv
   +253 Z^6+142
   Z^7\right)+2 X^2 (-1+Z)^5 Z \left(27+225 Z+68 Z^2-256 Z^3+119
   Z^4+197 Z^5\dvv
   +2 Z^6-6 Z^7+8 Z^8\right)-4 X^4 (-1+Z)^4 \left(-267+139
   Z-1071 Z^2-1008 Z^3+704 Z^4\dvv
   +99 Z^5-35 Z^6+226 Z^7+61
   Z^8\right)\Big]\, .
\end{align}

\begin{align}
\delta B&=-\frac{2\lambda_{1}}{\mathcal{D}} \Big[112640 X^{18}+1536 X^{16} \left(-267+212 Z+63 Z^2\right)\dv
+256
   X^{14} \left(2589-3962 Z
   +194 Z^2+1126 Z^3+141 Z^4\right)\dv
   +128 X^{12}
   \left(-4863+10460 Z-3828 Z^2-4210 Z^3+1835 Z^4+534 Z^5+72
   Z^6\right)\dv
   -32 X^{10} (-1+Z)^2 \left(-11313+7582 Z+6767 Z^2-5740
   Z^3-1615 Z^4+454 Z^5+297 Z^6\right)\dv
   -8 X^8 (-1+Z)^3 \left(-15819+2635
   Z+8769 Z^2-11265 Z^3-3925 Z^4+4181 Z^5+3535 Z^6\dvv
   +849
   Z^7\right)-(-1+Z)^7 Z^3 \left(54+81 Z+Z^2-39 Z^3-3 Z^4+15 Z^5+15
   Z^6+15 Z^7+5 Z^8\right)\dv
   -4 X^6 (-1+Z)^4 \left(-5973-1626 Z+1744
   Z^2-8694 Z^3-5370 Z^4+3426 Z^5+4696 Z^6\dvv
   +1822 Z^7+87 Z^8\right)+X^2
   (-1+Z)^6 Z \left(54+261 Z+208 Z^2-144 Z^3-324 Z^4-42 Z^5\dvv
   +336 Z^6+480
   Z^7+206 Z^8+21 Z^9\right)+2 X^4 (-1+Z)^5 \left(930+610 Z+333
   Z^2+3179 Z^3\dvv
   +2641 Z^4-889 Z^5-1749 Z^6-451 Z^7+453 Z^8+159
   Z^9\right)\Big]\, ,
\end{align}

\begin{align}
\delta\hat{A}_{1}&=\frac{\lambda_{1}}{\mathcal{D}}\Big[55296 X^{18}+512 X^{16} \left(-277+320 Z+53 Z^2\right)\dv
+256 X^{14}
   \left(543-1390 Z+414 Z^2+354 Z^3+55 Z^4\right) \dv
   +64
   X^{12} \left(-961+4158 Z-3017 Z^2-1432 Z^3+761 Z^4+346 Z^5+145
   Z^6\right)\dv
   -64 X^{10} (-1+Z)^2
   \left(-97+804 Z+1143 Z^2-204 Z^3-601 Z^4-124 Z^5+111 Z^6\right)\dv
  -8 X^8 (-1+Z)^3 \left(645+559 Z+8593 Z^2+3115 Z^3-5509
   Z^4-1567 Z^5+1663 Z^6\dvv
   +469 Z^7\right)-3 (-1+Z)^7 Z^3 \left(54+81
   Z+Z^2-39 Z^3-3 Z^4+15 Z^5+15 Z^6+15 Z^7\dvv
   +5 Z^8\right)+4 X^6 (-1+Z)^4
   \left(-423-298 Z-8344 Z^2-5354 Z^3+5854 Z^4+2442 Z^5\dvv-1544 Z^6-198
   Z^7+281 Z^8\right)+2 X^4 (-1+Z)^5 \left(-18-138 Z-2995 Z^2-2161
   Z^3\dvv
   +2563 Z^4-127 Z^5-1681 Z^6+693 Z^7+1187 Z^8+277 Z^9\right)+X^2
   (-1+Z)^6 Z \left(-54\dvv
   -27 Z+72 Z^2+250 Z^3-756 Z^4-908 Z^5+184 Z^6+550
   Z^7+202 Z^8+7 Z^9\right)\Big]\, ,
\end{align}
\begin{align}
\delta\hat{A}_{2}&=\frac{4\lambda_{1} X^2}{(1-2 X^2-Z)\mathcal{D}} \Big[56320 X^{18}+512 X^{16} \left(-483+373 Z+122
   Z^2\right)\dv
   +256 X^{14} \left(1834-2831 Z+168 Z^2+789 Z^3+84
   Z^4\right)\dv
   +128 X^{12} \left(-3940+8968 Z-3991 Z^2-3099 Z^3+1631
   Z^4+403 Z^5+28 Z^6\right)\dv
   -16 X^{10} (-1+Z)^2 \left(-20713+19650
   Z+9361 Z^2-12280 Z^3-2591 Z^4+590 Z^5\dvv
   +175 Z^6\right)-8 X^8 (-1+Z)^3
   \left(-16471+9646 Z+5647 Z^2-14740 Z^3-2013 Z^4+3910 Z^5\dvv
   +2077
   Z^6+504 Z^7\right)-4 X^6 (-1+Z)^4 \left(-7201+1949 Z-1834 Z^2-15013
   Z^3-3262 Z^4\dvv+5335 Z^5+4494 Z^6+1993 Z^7+275 Z^8\right)+2 X^4 (-1+Z)^5 \left(1353-252 Z+3201 Z^2\dvv
   +9152
   Z^3+3133 Z^4-2454 Z^5-2829 Z^6-1788 Z^7-274 Z^8+54 Z^9\right)\dv
   +X^2
   (-1+Z)^6 \left(54-27 Z+1461 Z^2+2629 Z^3+417 Z^4-817 Z^5-329 Z^6-149
   Z^7\dvv
   +281 Z^8+236 Z^9+36 Z^{10}\right)-(-1+Z)^7 Z^2
   \left(-162-189 Z+108 Z^2+161 Z^3-38 Z^4\dvv
   -99 Z^5-104 Z^6-45 Z^7+4
   Z^8+4 Z^9\right)\Big]\, ,
\end{align}
where the denominator reads
\begin{align}
\mathcal{D}=&(Z-1)^4 (Z+1)^2 \Big[96 X^{12}+48 X^{10} \left(Z^2+8 Z-5\right)\dv-16 X^8 \left(2
   Z^4-11 Z^3-33 Z^2+49 Z-13\right)\dv
   -8 X^6 \left(5 Z^6+13 Z^5-42 Z^4-38
   Z^3+113 Z^2-63 Z+12\right)\dv
   -2 X^4 (Z-1)^2 \left(9 Z^6+44 Z^5+75
   Z^4-70 Z^3-150 Z^2+32 Z-12\right)\dv
   -X^2 (Z-1)^3 Z^2 \left(Z^5+9 Z^4+25
   Z^3+35 Z^2-46 Z-72\right)
   \dv+(Z-1)^4 Z^5 \left(Z^3+2
   Z^2+Z+2\right)\Big]\,.
\end{align}
\subsection{Supersymmetric solution with $J_1 \neq J_2$} \label{appsec:unequalJ}
When the angular momenta are distinct, the corrections to the supersymmetric solution are given by
\begin{align}
\begin{split}
W_{0,0}=&-\frac{1+5 X-15 X^2+45 X^3}{1-9 X-27 X^2+27 X^3}\, ,\\
W_{1,0}=&\frac{-7-12
   X-162 X^2+540 X^3+405 X^4}{(1+3 X)^2 \left(1-12 X+9
   X^2\right)^2}\, ,\\
W_{1,1}=& \frac{-1-16 X+198 X^2-600 X^3+639 X^4-216
   X^5}{(-1+X) (1+3 X) \left(-1+3 X^2\right) \left(1-12 X+9
   X^2\right)}\, ,
\end{split}	\\ \nonumber \\ \begin{split}
G_{0,0}=&\frac{4 (-1+3 X) \left(-1+11 X-69 X^2+99 X^3\right)}{X (1+3 X) \left(1-12 X+9 X^2\right)}\, ,\\
G_{1,0}=&\frac{2 \left(-1+18 X+57 X^2-1476 X^3+7533 X^4-10854 X^5+243 X^6\right)}{X^2 (1+3 X)^2 \left(1-12 X+9 X^2\right)^2}\, ,\\
G_{1,1}=&\frac{-11+188 X-551 X^2-1248 X^3+6603 X^4-7740 X^5+2727 X^6}{2 (-1+X) X (1+3 X) \left(-1+3 X^2\right) \left(1-12 X+9 X^2\right)}\, ,
\end{split}
\end{align}

\begin{align}
\begin{split}
B_{0,0}=&-\frac{9 X \left(3+2 X-69 X^2+126 X^3\right)}{2 (1+3 X)^2 \left(1-12 X+9 X^2\right)}\, ,\\
B_{1,0}=&\frac{9 \left(-3+5 X+42 X^2-1782 X^3+4509 X^4+405 X^5\right)}{4 (1+3 X)^3 \left(1-12 X+9 X^2\right)^2}\, ,\\
B_{1,1}=&-\frac{3 \left(6+31 X-807 X^2+2958 X^3-3096 X^4-513 X^5+1701 X^6\right)}{4 (-1+X) (1+3 X)^2 \left(-1+3 X^2\right) \left(1-12 X+9 X^2\right)}\, ,
\end{split}	\\ \nonumber \\ \begin{split}
J_{0,0}=&-\frac{4 \eta (1+3 X)^2 \left(-9+32 X-177 X^2+180 X^3\right)}{9 X^4 \left(1-12 X+9 X^2\right)}\, ,\\
J_{1,0}=&-\frac{2 (1+3 X) \left(63-1132 X+3801 X^2-12024 X^3+20709 X^4-19116
   X^5+9963 X^6\right)}{9 X^4 \eta \left(1-12 X+9
   X^2\right)^2}\, ,\\
J_{1,1}=&-\frac{\eta (1+3 X)^2 \left(36-133 X+1605 X^2-6234
   X^3+13446 X^4-14661 X^5+2997 X^6\right)}{27 (-1+X) X^5 \left(-1+3
   X^2\right) \left(1-12 X+9 X^2\right)}\, ,
\end{split}	\\ \nonumber \\ \begin{split}
\hat{A}_{1,0,0}=&\frac{4 (1+3 X) \left(9-59 X+273 X^2-711 X^3+540 X^4\right)}{3 \sqrt{3}
   X^3 \left(1-12 X+9 X^2\right)}\, ,\\
\hat{A}_{1,1,0}=&-\frac{2 \left(27-496 X+1653 X^2-3456 X^3+2889 X^4-1296 X^5+5103
   X^6\right)}{3 \sqrt{3} X^4 \left(1-12 X+9 X^2\right)^2}\, ,\\
\hat{A}_{1,1,1}=&-\frac{(1+3 X)}{9 \sqrt{3} (-1+X) X^4 \left(-1+3
   X^2\right) \left(1-12 X+9 X^2\right)} \\ &\times \left(36-217 X+1968 X^2-5559 X^3-3708 X^4+31293
   X^5-36936 X^6+13851 X^7\right)\, ,
\end{split}	\\ \nonumber \\ \begin{split}
\hat{A}_{2,0,0}=&\frac{3 \sqrt{3} \eta \left(5-36 X+117 X^2\right)}{2 (1+3
   X) \left(1-12 X+9 X^2\right)}\, ,\\
\hat{A}_{2,1,0}=&\frac{3 \sqrt{3} \left(5-78 X+1530 X^2-7236 X^3+6885 X^4+3402
   X^5\right)}{4 \eta (1+3 X)^2 \left(1-12 X+9
   X^2\right)^2}\, ,\\
\hat{A}_{2,1,1}=&\frac{\sqrt{3} \eta \left(-4-51 X+924 X^2-2754 X^3+2160
   X^4+81 X^5\right)}{4 (-1+X) X (1+3 X) \left(-1+3 X^2\right)
   \left(1-12 X+9 X^2\right)}\, ,
   \end{split}
\end{align}
where
\begin{align}
	\eta = \sqrt{X(3X-2)}\,.
\end{align}

\bibliographystyle{JHEP}
\bibliography{Gravities.bib}

\end{document}